\documentclass[11pt,a4paper]{article}

\usepackage{fullpage}
\usepackage{microtype}
\usepackage{xcolor}
\usepackage{tikz}
\usepackage[OT1]{fontenc}
\usepackage[tt=false]{libertine}
\usepackage{amsmath}
\usepackage{amssymb}
\usepackage{amsthm}
\usepackage{mathtools}
\usepackage{enumitem}
\usepackage{float}
\usepackage{subcaption}
\usepackage{csquotes}
\usepackage[
  backend=biber,
  style=alphabetic,
  url=false,
  doi=true,
  isbn=false,
  maxalphanames=4,
  minalphanames=3,
  maxnames=99,
  backref=false,
]{biblatex}

\definecolor{CitationBurgundy}{HTML}{8B3A4A}
\definecolor{ReferenceBlue}{HTML}{2B5C8A}
\definecolor{ExternalTeal}{HTML}{256D63}
\usetikzlibrary{arrows.meta}

\definecolor{cink}{HTML}{1D2433}
\definecolor{cmuted}{HTML}{5F6675}
\definecolor{chair}{HTML}{D6CFBF}
\definecolor{cedge}{HTML}{8E96A3}
\definecolor{cterra}{HTML}{B9472F}

\definecolor{gx}{HTML}{C8553D}
\definecolor{gblue}{HTML}{3F7CB5}
\definecolor{ggold}{HTML}{E0A43A}
\definecolor{gteal}{HTML}{2F9A87}
\definecolor{gpurple}{HTML}{7D5BA6}

\tikzset{
  vert/.style={circle,draw=cink,line width=0.5pt,minimum size=0.30cm,inner sep=0pt},
  ring/.style={circle,draw=gx,line width=1.5pt,minimum size=0.47cm,inner sep=0pt},
  gedge/.style={cedge,line width=0.7pt},
  move/.style={-{Stealth[length=5pt,width=4pt]},cink,line width=0.8pt},
  nomove/.style={cedge,line width=0.7pt,dash pattern=on 2.5pt off 2pt},
  flab/.style={font=\footnotesize,text=cmuted,inner sep=1.5pt},
  ftitle/.style={font=\small},
  fcap/.style={font=\footnotesize,text=cink},
  pics/pair/.style 2 args={code={
    \draw[cink,line width=0.9pt] (-0.36,0) -- (0.36,0);
    \node[vert,fill=#1] at (-0.36,0) {};
    \node[vert,fill=#2] at (0.36,0) {};
  }},
}

\newcommand*{\rectcorner}[2]{\tikz[x=0.85cm,y=0.55cm,baseline={([yshift=-0.5ex]current bounding box.center)}]{\draw[chair,line width=0.8pt] (0,0) rectangle (1,-1);
    \draw[cterra,line width=1.4pt,line cap=round,line join=round] ({1-#1},{-#2}) -- (#1,{-#2}) -- (#1,{#2-1});
    \fill[cterra] (#1,{-#2}) circle (1.9pt);}}

\renewbibmacro{in:}{}
\renewcommand*{\bibpagerefpunct}{\addperiod\space}
\renewbibmacro*{pageref}{\iflistundef{pageref}
  {}
  {\printlist[pageref][-\value{listtotal}]{pageref}\nopunct}}
\renewbibmacro*{setpageref}{\iflistundef{pageref}
  {}
  {\setunit{\bibpagerefpunct}\newblock
    \printlist[pageref][-\value{listtotal}]{pageref}\nopunct}}
\usepackage{ifthen}
\usepackage{xifthen}
\usepackage[bookmarks=true,hypertexnames=false]{hyperref}
\hypersetup{
  colorlinks=true,
  citecolor=CitationBurgundy,
  linkcolor=ReferenceBlue,
  urlcolor=ExternalTeal,
}
\usepackage{xurl}
\usepackage{prettyref}

\newrefformat{thm}{\hyperref[#1]{Theorem~\ref*{#1}}}
\newrefformat{lem}{\hyperref[#1]{Lemma~\ref*{#1}}}
\newrefformat{pps}{\hyperref[#1]{Proposition~\ref*{#1}}}
\newrefformat{def}{\hyperref[#1]{Definition~\ref*{#1}}}
\newrefformat{cor}{\hyperref[#1]{Corollary~\ref*{#1}}}
\newrefformat{fact}{\hyperref[#1]{Fact~\ref*{#1}}}
\newrefformat{eq}{\hyperref[#1]{(\ref*{#1})}}
\newrefformat{sec}{\hyperref[#1]{Section~\ref*{#1}}}
\newrefformat{app}{\hyperref[#1]{Appendix~\ref*{#1}}}
\newrefformat{fig}{\hyperref[#1]{Figure~\ref*{#1}}}
\newcommand{\pref}{\prettyref}

\def\*#1{\mathbf{#1}}
\def\+#1{\mathcal{#1}}
\def\-#1{\mathrm{#1}}
\def\^#1{\mathbb{#1}}
\def\!#1{\mathfrak{#1}}
\def\$#1{\mathscr{#1}}

\newcommand{\defeq}{\coloneqq}

\newtheorem{theorem}{Theorem}

\newtheorem{lemma}[theorem]{Lemma}

\newtheorem{corollary}[theorem]{Corollary}
\newtheorem{fact}[theorem]{Fact}
\newtheorem*{conjecture*}{Conjecture}
\theoremstyle{definition}

\theoremstyle{remark}
\newtheorem*{remark}{Remark}

\newcommand{\tuple}[1]{\mathopen{}\mathclose{\left(#1\right)}}
\newcommand{\set}[2][]{\ifthenelse{\isempty{#1}}
  {\left\{#2\right\}}
  {\left\{#2\right\}_{#1}}}
\newcommand{\one}{\mathbf 1}

\DeclareMathOperator{\oPr}{\mathbb{P}}
\renewcommand{\Pr}[2][]{\ifthenelse{\isempty{#1}}
  {\oPr\left[#2\right]}
  {\oPr_{#1}\left[#2\right]}}

\DeclareMathOperator*{\oE}{\mathbb{E}}
\newcommand{\E}[2][]{\ifthenelse{\isempty{#1}}
  {\oE\left[#2\right]}
  {\oE_{#1}\left[#2\right]}}

\DeclareMathOperator*{\oVar}{Var}
\newcommand{\Var}[2][]{\ifthenelse{\isempty{#1}}
  {\oVar\left[#2\right]}
  {\oVar_{#1}\left[#2\right]}}

\newcommand{\Id}{\mathrm{Id}}
\newcommand{\ee}{\mathrm{e}}

\DeclareMathOperator{\gap}{gap}
\newcommand{\ip}[3][\mu]{{\langle #2,#3\rangle}\if\relax\detokenize{#1}\relax\else_{#1}\fi}
\newcommand{\norm}[2][\mu]{{\lVert #2\rVert}\if\relax\detokenize{#1}\relax\else_{#1}\fi}
\newcommand{\muh}{\widehat{\mu}}
\newcommand{\lift}[1]{{#1}^{\uparrow}}
\newcommand{\down}[1]{{#1}^{\downarrow}}
\newcommand{\tmix}{t_{\mathrm{mix}}}
\providecommand{\dTV}{d_{\mathrm{TV}}}

\usepackage{titling}
\title{Glauber Dynamics for Sampling Colorings:\\ Rapid Mixing above $(1+1/\mathrm{e})\Delta$ via Hidden Hard-Core Models}
\date{}

\author{Xiongxin Yang\\[0.1em] University of California, Santa Barbara \\[0.1em] \texttt{xiongxinyang@ucsb.edu}}

\begin{document}

\maketitle

\begin{abstract}
We show that for every $\gamma>0$ and every sufficiently large $\Delta$, the Glauber dynamics for proper $q$-colorings of any $n$-vertex graph of maximum degree $\Delta$ has spectral gap $\Omega_\gamma(1/n)$, and hence mixing time $O_\gamma(n^2\log q)$, whenever $q\ge(1+1/\mathrm{e}+\gamma)\Delta$. 
Our proof establishes a Bochner-type inequality by exploiting the spectral gaps of hard-core models hidden in the color classes. 
The constant $1+1/\mathrm{e}\approx1.368$ comes from the uniqueness threshold $\approx\mathrm{e}/\Delta$ of the hard-core model on the infinite $\Delta$-regular tree.
\end{abstract}

\section{Introduction}\label{sec:intro}

Sampling a uniformly random proper $q$-coloring of a graph $G$ is a fundamental task in theoretical computer science, discrete mathematics, and statistical physics,
yet determining the minimum number of colors required for efficient sampling remains a major challenge.
Perhaps the simplest and most widely studied sampler is the (heat-bath) \emph{Glauber dynamics},
which repeatedly picks a uniformly random vertex and recolors it with a uniformly random color absent from its neighborhood.
A long-standing conjecture asserts that this dynamics mixes rapidly whenever $q\ge\Delta+2$.

\begin{conjecture*}[\cite{Jer95Very}]
  For any graph $G$ with maximum degree $\Delta$, if $q\ge\Delta+2$, then the Glauber dynamics mixes in time polynomial in $n$.
\end{conjecture*}

Rapid mixing for $q>2\Delta$ was established using coupling arguments by Jerrum~\cite{Jer95Very} and by Salas and Sokal~\cite{SS97Absence}.
Vigoda~\cite{Vig00Improved} improved this bound to $q>11\Delta/6$ by analyzing the \emph{flip dynamics},
using a path-coupling analysis to establish rapid mixing and a comparison argument to transfer this result to Glauber dynamics.
After standing for nearly two decades as the best bound on general graphs,
the $11/6$ barrier was broken by Chen, Delcourt, Moitra, Perarnau, and Postle~\cite{CDMPP19Improved},
who achieved $q\ge(11/6-\varepsilon_0)\Delta$ for a small absolute constant $\varepsilon_0>0$ by combining linear programming with refined coupling arguments for the flip dynamics.
Carlson and Vigoda~\cite{CV25Flip} subsequently reached $q\ge1.809\Delta$ for $\Delta\ge125$ using path coupling with a weighted Hamming metric.
Very recently, Li and Oveis Gharan~\cite{LO26Local} pushed the threshold down to $q\ge1.7408\Delta$ for sufficiently large $\Delta$ by establishing a Bochner-type inequality for the Glauber dynamics,
which yields a spectral gap of $\Omega(1/n)$.

In this paper, we lower the threshold to $(1+1/\ee+\gamma)\Delta$ colors (where $1+1/\ee\approx1.368$) for every fixed $\gamma>0$ and all sufficiently large $\Delta$,
likewise by establishing a Bochner-type inequality.

\begin{theorem}\label{thm:main}
  For every $\gamma>0$, there exist $\Delta_0(\gamma)<\infty$ and $\kappa_\gamma>0$ such that for every $n$-vertex graph $G$ of maximum degree at most $\Delta\ge\Delta_0(\gamma)$ and every $q\ge(1+1/\ee+\gamma)\Delta$,
  the Glauber dynamics $P$ for proper $q$-colorings of $G$ has spectral gap at least $\kappa_\gamma/n$, and hence satisfies, for every $\varepsilon\in(0,1)$,
  \begin{equation*}
    \tmix\tuple{P,\varepsilon}\le\frac n{\kappa_\gamma}\tuple{n\log q+\log\frac1\varepsilon}.
  \end{equation*}
\end{theorem}
Since $\Delta_0(\gamma)$ and $\kappa_\gamma$ depend only on $\gamma$, the bounds hold uniformly even when $\Delta$ and $q$ grow with $n$.

\subsection*{Related work}

\paragraph{Sampling colorings with structural assumptions}

Local sparsity allows fewer colors.
Let $\alpha_0\approx1.763$ be the positive root of $\alpha_0=\ee^{1/\alpha_0}$.
Burn-in coupling arguments exploiting local uniformity~\cite{DF03Randomly,Hay03Randomly,Mol04Glauber,FV06Randomly} progressively relaxed girth and color assumptions under the condition $\Delta=\Omega(\log n)$.
Dyer, Frieze, Hayes, and Vigoda~\cite{DFHV13Randomly} removed this degree condition,
proving, for sufficiently large $\Delta$, $O_\varepsilon(n\log n)$ mixing for $q\ge(\alpha_0+\varepsilon)\Delta$ at girth at least~$5$ and for $q\ge1.4891\Delta$ at girth at least~$7$.
For triangle-free graphs with $q\ge(\alpha_0+\varepsilon)\Delta$,
spectral independence yielded polynomial mixing for arbitrary $\Delta$~\cite{CGSV21Rapid,FGYZ22Rapid,CLV21Optimal,JPV22Spectral},
and Chen and Feng~\cite{CF25Coupling} finally obtained an optimal $\Omega_\varepsilon(1/n)$ spectral gap uniformly over $\Delta\ge3$.

Under stronger assumptions, the threshold approaches $\Delta$.
For fixed $\delta>0$ and $q\ge(1+\delta)\Delta$, Hayes and Vigoda~\cite{HV03NonMarkovian} introduced a non-Markovian coupling on large-girth graphs with $\Delta=\Omega_\delta(\log n)$.
Building on this approach for large $\Delta$,
Jain, Mizgerd, and Vigoda~\cite{JMV26Sampling} proved $O_\delta(n\log n)$ mixing (for the Metropolis Glauber dynamics) at girth at least~$7$,
while Chen and Liu~\cite{CL26Girth} established an $\Omega_\delta(1/n)$ spectral gap at girth at least~$5$.
At girth sufficiently large in $\Delta$, Chen, Liu, Mani, and Moitra~\cite{CLMM23Strong} proved $O_\Delta(n\log n)$ mixing for $q\ge\Delta+3$.
On line graphs, Wang, Zhang, and Zhang~\cite{WZZ24Sampling} used matrix trickle-down to prove $O_\Delta(n\log n)$ mixing of the Glauber dynamics for $q>(1+o(1))\Delta$.
Haqi~\cite{Haq26Sampling} subsequently established rapid mixing for $q\ge\Delta+5$ on line graphs.

Planar graphs and trees require even fewer colors.
Polynomial mixing holds on planar graphs for $q=\Omega(\Delta/\log\Delta)$ at large $\Delta$~\cite{HVV15Randomly}.
On trees with free boundary, the mixing time is at most $n^{O(1+\Delta/(q\log\Delta))}$ for $q\ge4$ and $\Delta\ge3$~\cite{LM11Glauber}.
On complete $b$-ary trees, the mixing time undergoes a phase transition at $q=(1+o_b(1))b/\log b$,
whose location agrees to first order with the reconstruction threshold~\cite{TVVY12Phase}.

Finally, on sparse random graphs $G(n,d/n)$, thresholds scale with the average degree $d$.
With high probability for fixed $\varepsilon\in(0,1)$ and sufficiently large constant $d$,
polynomial mixing was first shown for $q\ge C(d)$~\cite{MS10Gibbs} and later improved to $q\ge(\alpha_0+\varepsilon)d$~\cite{EHSV18Sampling} and $q\ge(1+\varepsilon)d$~\cite{LY26Rapid}.

\paragraph{Bochner-type inequalities}

Our proof relies on an integrated Bochner inequality.
Tracing back to Bochner~\cite{Boc46Vector} and long used to bound spectral gaps and entropy decay in interacting particle systems~\cite{BCDP06Spectral,CDP09Convex,KKO13Spectral},
Bochner-type identities have recently found wide application in approximate counting and sampling.
Beyond the coloring results~\cite{CL26Girth,Haq26Sampling,LO26Local} discussed above,
examples include a short proof of rapid mixing for the hard-core model on random regular graphs beyond the uniqueness threshold~\cite{GJMPPS26Simple} (first established via trickle-down~\cite{CCCYZ25Rapid}),
a unified Garland-style~\cite{Gar73Padic} treatment of trickle-down theorems~\cite{CL26Trickle},
and a related operator expansion for the planar hard-core model at small constant activity~\cite{GZ26Approximating}.
Further recent applications establish spectral-gap and mixing bounds for the Sherrington--Kirkpatrick model~\cite{Wan26Optimal,BLO26Rank},
common bases of two matroids~\cite{CL26Common}, and the permanent~\cite{CGVY26Permanent}.

\section{Overview of the proof}

Set $L\defeq n(\Id-P)$.
By the spectral theorem, a lower bound of order $1/n$ on the spectral gap is equivalent to the \emph{integrated Bochner inequality} (\pref{lem:spectral})
\begin{equation*}
  \norm{Lf}^2\ge\kappa_\gamma\ip f{Lf}\qquad\text{for every $f\in L^2(\mu)$},
\end{equation*}
which is the integrated form of the $\Gamma_2$-criterion in the language of Bakry and \'Emery~\cite{BE85Diffusions,BGL14Analysis}.
We prove it by bounding a sum of single-color terms in two ways.
A local comparison bounds this sum from above by $\norm{Lf}^2$, and a global comparison, based on hard-core models hidden in the color classes,
bounds it from below by $\ip f{Lf}$.

\paragraph{The local comparison.}
Write $L=\sum_vQ_v$, where $Q_v=\Id-P_v$ subtracts from each function its conditional expectation given $\sigma_{V\setminus\{v\}}$.
Expanding $\norm{Lf}^2$ gives the Dirichlet form $\ip f{Lf}$ plus cross terms indexed by pairs of vertices.
Updates at non-adjacent vertices commute, so their cross terms are nonnegative.
For an edge $uv$, the cross term is a weighted sum over pairs of individually legal recolorings, one at each endpoint.
If the two new colors differ, the recolorings can also be performed together, and the four resulting colorings form a rectangle,
as observed in~\cite{Haq26Sampling}.
Thus a four-corner identity groups the corresponding terms into squares, up to a small error caused by the heat-bath weights.
If the two new colors coincide, the fourth coloring is not proper, and these terms cannot be grouped into squares.

The latter terms are the cross terms of the \emph{single-color operators}.
For a color $c$, let $L_c$ retain the moves of $L$ whose old or new color is $c$.
At a fixed vertex, these moves form a star centered at $c$ among the available colors,
and every nontrivial recoloring belongs to the stars of its old and new colors, so $\sum_cL_c=2L$.
Comparing the expansions of $\norm{Lf}^2$ and $\sum_c\norm{L_cf}^2$ term by term yields
\begin{equation*}
  \sum_{c\in[q]}\norm{L_cf}^2\le\tuple{1+\frac2{q-\Delta}}\norm{Lf}^2+\frac{50\Delta}{9\tuple{q-\Delta}^2}\ip f{Lf}.
\end{equation*}
See \pref{lem:bochner-comparison}.
Both error coefficients are of order $O(1/\Delta)$ whenever $q-\Delta$ is at least a fixed multiple of $\Delta$.

\paragraph{Hidden hard-core models.}
Fix a color $c$, and let $I$ be its color class in a uniformly random coloring.
Recolor each vertex of $I$ independently with a uniformly random \emph{ghost color}, chosen among the colors other than $c$ available to it.
The result is a \emph{background coloring} $\eta$, a proper coloring in which no vertex has color $c$;
the original coloring is recovered by assigning color $c$ to every vertex in $I$.
Conditionally on $\eta$, the set $I$ follows a hard-core model $\nu_\eta$ with activities
\begin{equation*}
  \lambda_v\tuple{\eta}=\frac1{|A_v\tuple{\eta}|-1}\le\frac1{q-\Delta-1}.
\end{equation*}
See \pref{lem:lift}.
The \emph{hidden dynamics} is the Glauber dynamics for $\nu_\eta$;
it updates $I$ while $\eta$ stays fixed, and we write $H_\eta$ for its operator, with the same normalization as $L$.
The \emph{uniqueness threshold} of the hard-core model on the infinite $\Delta$-regular tree is asymptotically $\ee/\Delta$,
so for $q\ge(1+1/\ee+\gamma)\Delta$ and $\Delta$ sufficiently large, all hidden models lie uniformly inside the uniqueness regime;
this is the origin of the constant $1+1/\ee$.
The spectral gap theorem of Chen, Feng, Yin, and Zhang~\cite{CFYZ21Rapid} then gives $\gap(H_\eta)\ge\theta_\gamma$ for a constant $\theta_\gamma>0$,
uniformly in $c$, $\eta$, $\Delta$, and the graph.

\paragraph{Mean and variance parts.}
View $f$ as a function of $(\eta,I)$ through the coloring obtained by assigning color $c$ to every vertex in $I$.
Taking the expectation of the Dirichlet form of the hidden dynamics over the ghost colors then recovers $\ip f{L_cf}$ (\pref{lem:compression}).
However, the Poincar\'e inequality for each hidden model controls only the variance of the lifted function conditional on $\eta$.
Taking the expectation over $\eta$ therefore bounds the expectation of these conditional variances,
rather than the full variance required for a spectral-gap bound on $L_c$.
The variation of the conditional expectation across backgrounds remains uncontrolled.
Even for a single vertex, the hidden dynamics has spectral gap $1$, whereas $L_c$ has eigenvalue $1/q$.

We therefore separate two kinds of variation at a vertex at which $c$ is available.
The \emph{mean part} $\+M_c(f)$ measures the difference between the value of $f$ at color $c$ and its average over the other available colors,
and the \emph{variance part} $\+V_c(f)$ measures the variation among those other values;
both carry heat-bath weights and are summed over all vertices.
Equivalently, $\+M_c(f)$ is the sum over vertices of the squared $L^2(\mu)$ norms of the conditional rank-one projections of $f$ associated with color $c$ used by Li and Oveis Gharan~\cite{LO26Local}.
The two parts satisfy
\begin{equation*}
  \ip f{L_cf}=\+M_c\tuple{f}+\+V_c\tuple{f},\qquad
  \sum_{c\in[q]}\+M_c\tuple{f}\ge\ip f{Lf},\qquad
  \sum_{c\in[q]}\+V_c\tuple{f}\le\ip f{Lf}.
\end{equation*}
See \pref{lem:MV-split} and \pref{lem:mean-sum}.
So after summing over colors, the mean parts alone control the Dirichlet form.
The mean parts are also compatible with averaging over the ghost colors.
By Jensen's inequality, for every function of $(\eta,I)$, the expectation of the Dirichlet form of the hidden dynamics is at least the mean part of its ghost-color average (\pref{lem:compression}).

\paragraph{Smoothing by resolvents.}
For each color $c$, we compare $f$ with a smoothed function $\widetilde f_c$ whose mean part is small.
Polarizing the mean--variance decomposition gives the exact identity
\begin{equation*}
  \+M_c\tuple{f}=\ip{f-\widetilde f_c}{L_cf}-\+V_c\tuple{f-\widetilde f_c,f}+\+M_c\tuple{\widetilde f_c,f}.
\end{equation*}
For any $a>0$, the Cauchy--Schwarz and Young inequalities, together with $\sum_c\+V_c(f)\le\ip f{Lf}$, give
\begin{align*}
  \sum_{c\in[q]}\ip{f-\widetilde f_c}{L_cf}
  &\le\frac1{8a}\sum_{c\in[q]}\norm{f-\widetilde f_c}^2
  +2a\sum_{c\in[q]}\norm{L_cf}^2,\\
  -\sum_{c\in[q]}\+V_c\tuple{f-\widetilde f_c,f}
  &\le\frac18\ip f{Lf}+2\sum_{c\in[q]}\+V_c\tuple{f-\widetilde f_c},\\
  \sum_{c\in[q]}\+M_c\tuple{\widetilde f_c,f}
  &\le\frac14\sum_{c\in[q]}\+M_c\tuple{f}
  +\sum_{c\in[q]}\+M_c\tuple{\widetilde f_c}.
\end{align*}
We therefore need to control $\sum_c\norm{f-\widetilde f_c}^2$, make $\sum_c\+V_c(f-\widetilde f_c)$ small,
and bound $\sum_c\+M_c(\widetilde f_c)$ by a sufficiently small fraction of $\ip f{Lf}$.
Since $\sum_c\+M_c(f)\ge\ip f{Lf}$, the third bound then allows us to absorb the mean cross term into the LHS.

To control the displacement while reducing the mean part, for each fixed background coloring $\eta$ and $h\in L^2(\nu_\eta)$ we minimize
\begin{equation*}
  \norm[\nu_\eta]{g-h}^2+a\ip[\nu_\eta]g{H_\eta g}
  \qquad\text{over }g\in L^2\tuple{\nu_\eta}.
\end{equation*}
The first term penalizes displacement from $h$, while the second penalizes variation along the hidden updates.
The unique minimizer is $R_\eta h$, where $R_\eta\defeq(\Id+aH_\eta)^{-1}$ is the \emph{resolvent} of the hidden dynamics.
We therefore view $f$ as a function of $(\eta,I)$, apply this minimization separately for each $\eta$,
and take the expectation over the ghost colors to obtain $\widetilde f_c$.
Moreover, $R_\eta$ multiplies each eigenfunction of $H_\eta$ with eigenvalue $z$ by $1/(1+az)$.
Since every nonconstant eigenfunction has $z\ge\theta_\gamma$, the Dirichlet form contracts by the factor $(1+a\theta_\gamma)^{-2}$.
As shown in \pref{lem:mean}, taking the expectation over the ghost colors and summing over $c$ then gives
\begin{equation*}
  \sum_{c\in[q]}\norm{f-\widetilde f_c}^2\le2a\ip f{Lf},\qquad
  \sum_{c\in[q]}\+M_c\tuple{\widetilde f_c}\le\frac2{\tuple{1+a\theta_\gamma}^2}\ip f{Lf}.
\end{equation*}
Taking $a$ to be a sufficiently large constant multiple of $1/\theta_\gamma$ therefore makes the sum of the smoothed mean parts a fixed small fraction of $\ip f{Lf}$.

\paragraph{Background recolorings.}
It remains to bound the variance part of the displacement, which measures the sensitivity of $f-\widetilde f_c$ to recolorings among the non-$c$ colors.
We express this sensitivity through background recolorings on the lifted space.
Changing the background coloring at one vertex changes both the lifted function and the hidden dynamics, and the resolvent identity
\begin{equation*}
  R_{\eta'}-R_\eta=aR_{\eta'}\tuple{H_\eta-H_{\eta'}}R_\eta
\end{equation*}
separates the two effects.
If $\eta$ and $\eta'$ differ at a single vertex, the activities differ only at its neighbors, where the number of available colors changes by at most $1$.
The change of the lifted function, combined with a hidden update at another vertex,
produces mixed second differences of $f$ with the same four-corner structure as in the local comparison.
Moreover, when expressed through background recolorings, the variance part at a vertex $v$ carries the factor $p_v(\eta)\le1/(q-\Delta)$,
the probability of a birth at $v$.
Together, these estimates give
\begin{equation*}
  \sum_{c\in[q]}\+V_c\tuple{f-\widetilde f_c}
  \le\frac{C_0a}{q-\Delta}\tuple{\norm{Lf}^2+\ip f{Lf}}+\frac{C_0a^2}{\tuple{q-\Delta}^2}\ip f{Lf},
\end{equation*}
an error of order $O_\gamma(1/\Delta)$ relative to $\norm{Lf}^2+\ip f{Lf}$.
See \pref{lem:variance} and \pref{lem:rect}.

\paragraph{Completing the proof.}
With $a=3/\theta_\gamma$, combining the bounds above gives
\begin{equation*}
  \sum_{c\in[q]}\norm{L_cf}^2\ge\frac1{8a}\ip f{Lf}-\frac1a\sum_{c\in[q]}\+V_c\tuple{f-\widetilde f_c},
\end{equation*}
and the variance estimate turns this into the global comparison (\pref{lem:hardcore-comparison})
\begin{equation*}
  \sum_{c\in[q]}\norm{L_cf}^2\ge\frac{\theta_\gamma}{24}\ip f{Lf}-O_\gamma\tuple{\frac1\Delta}\tuple{\norm{Lf}^2+\ip f{Lf}}.
\end{equation*}
Combined with the local comparison, this yields the integrated Bochner inequality once $\Delta$ is sufficiently large in terms of $\gamma$.
We state both comparisons and deduce \pref{thm:main} in \pref{sec:comparisons};
the local comparison is proved in \pref{sec:bochner}, and the global comparison in \pref{sec:hidden} and~\pref{sec:resolvent}.

\begin{remark}[Spectral independence]
  Our proof does not use spectral independence~\cite{ALO20Spectral}, but it implies it.
  Let $\mu^\tau$ be the uniform distribution on the colorings that extend a feasible pinning $\tau$, and let $U$ be the set of unpinned vertices.
  The proof applies verbatim to $\mu^\tau$, with hard-core models on induced subgraphs of $G$ as hidden models, 
  so its Glauber dynamics has spectral gap at least $\kappa_\gamma/|U|$.
  As in~\cite{AJKPV24Universality}, applying the Poincar\'e inequality to additive functions $f(\sigma)=\sum_{v}g_v(\sigma_v)$ 
  gives $\Var[\mu^\tau]{f}\le\kappa_\gamma^{-1}\sum_{v}\E[\mu^\tau]{g_v(\sigma_v)^2}$,
  which bounds the largest eigenvalue of the influence matrix by $\kappa_\gamma^{-1}$.
  Since this holds for any pinning, $\mu$ is $O_\gamma(1)$-spectrally independent.
  Moreover, every positive single-site marginal probability of $\mu^\tau$ is at least $\Omega(1/q)$, 
  so \cite{CLV21Optimal} gives $O_{\gamma,\Delta,q}(n\log n)$ mixing time for bounded $\Delta$ and $q$.
\end{remark}

\paragraph{Comparison with concurrent works.}
Several recent works also analyze the Glauber dynamics for colorings via Bochner-type inequalities.
Here we briefly compare their approaches to ours.

Chen and Liu~\cite{CL26Girth} prove an $\Omega_\delta(1/n)$ spectral gap for $q\ge(1+\delta)\Delta$ on graphs of girth at least~$5$, 
for fixed $\delta\in(0,1)$ and sufficiently large $\Delta$.
Establishing a spectral local-to-global principle, 
they localize the cross terms of $\norm{Lf}^2$ by vertex and reduce the global Bochner inequality to spectral-gap estimates for weighted Glauber dynamics on stars 
(with exterior colors fixed), which they bound via Fourier analysis.
This reduction relies on the girth assumption, 
which ensures that each neighborhood is an independent set and that every pair of vertices at distance two shares a unique common neighbor.

Haqi~\cite{Haq26Sampling} obtains rapid mixing for $q\ge\Delta+5$ on line graphs.
For a dynamics in which every valid recoloring has rate~$1$, 
the cross terms split into legal rectangles and pairs of transitions sharing a new color.
Whereas Haqi controls the latter using the bound $\lambda_{\min}(A_G)\ge-2$ on line graphs, 
$\lambda_{\min}(A_G)$ can be as small as $-\Delta$ on general graphs; 
we therefore use the same split but retain the common-color terms inside the single-color operators.

Li and Oveis Gharan~\cite{LO26Local} establish an $\Omega_\alpha(1/n)$ spectral gap for list colorings with list sizes at least $q>\alpha\Delta$, 
for fixed $\alpha>\alpha^*\approx1.7408$ and sufficiently large $\Delta$.
Like us, they organize the cross terms by color, but they localize the expansion as in~\cite{CL26Girth} without a girth
assumption by averaging Bochner-type inequalities over a suitable distribution of cones 
(each consisting of a vertex and a subset of its neighbors).
Their threshold $\alpha^*$ arises from lower-bounding the probability that the leaves of a cone avoid a given color.

\section{Preliminaries}\label{sec:prelim}

\subsection{Markov chains and heat-bath updates}\label{sec:prelim-gap}

We use standard notions for finite discrete-time Markov chains;
see~\cite{LP17Markov}.
All Markov chains in this paper are reversible with respect to a distribution $\mu$ on a finite state space $\Omega$ with $\mu(\sigma)>0$ for every $\sigma\in\Omega$,
meaning that their transition matrices $P$ satisfy $\mu(\sigma)P(\sigma,\tau)=\mu(\tau)P(\tau,\sigma)$ for all $\sigma,\tau\in\Omega$;
in particular, $\mu$ is the stationary distribution for $P$.
Let $L^2(\mu)$ be the space of real-valued functions on $\Omega$,
equipped with the inner product $\ip fg\defeq\E[\sigma\sim\mu]{fg}$ and norm $\norm f^2\defeq\ip ff$,
where $\E[\sigma\sim\mu]{f}\defeq\sum_{\sigma\in\Omega}\mu(\sigma)f(\sigma)$.
The matrix $P$ acts on $L^2(\mu)$ by $(Pf)(\sigma)\defeq\sum_{\tau\in\Omega}P(\sigma,\tau)f(\tau)$ and is self-adjoint by reversibility.
We list its eigenvalues with multiplicity as $1=\lambda_1(P)\ge\lambda_2(P)\ge\cdots\ge\lambda_{|\Omega|}(P)\ge-1$ and define the \emph{spectral gap} $\gap(P)\defeq1-\lambda_2(P)$.
Write $P^t(\sigma,\cdot)$ for the distribution of the chain after $t$ steps started from $\sigma$,
and $\dTV(\nu,\mu)\defeq\frac12\sum_{\sigma\in\Omega}|\nu(\sigma)-\mu(\sigma)|$ for the total variation distance.
For $\varepsilon\in(0,1)$, the \emph{mixing time} of $P$ is
\begin{equation*}
  \tmix\tuple{P,\varepsilon}\defeq\min\set{t\ge0:\ \max_{\sigma\in\Omega}\dTV\tuple{P^t\tuple{\sigma,\cdot},\mu}\le\varepsilon}.
\end{equation*}
The following bound is standard;
see~\cite[Chapter~12]{LP17Markov}.

\begin{lemma}\label{lem:mixing}
  Let $\mu_{\min}\defeq\min_{\sigma\in\Omega}\mu(\sigma)$.
  If $P$ is irreducible and has no negative eigenvalues, then for every $\varepsilon\in(0,1)$,
  \begin{equation}\label{eq:mixing}
    \tmix\tuple{P,\varepsilon}\le\frac1{\gap\tuple{P}}\log\frac1{\varepsilon\mu_{\min}}.
  \end{equation}
\end{lemma}

\paragraph{Heat-bath updates.}
Suppose now that $\Omega\subseteq S^V$ for a finite set $V$ of $n$ vertices and a finite set $S$, and let $\mu$ be a distribution with full support on $\Omega$.
For $v\in V$, the \emph{heat-bath update} at $v$ resamples the coordinate $\sigma_v$ from its conditional distribution under $\mu$ given the configuration $\sigma_{V\setminus\{v\}}$ off $v$;
we write $P^\mu_v$ for its transition matrix.
The heat-bath \emph{Glauber dynamics} for $\mu$ updates a uniformly random vertex at each step,
so its transition matrix is $P^\mu\defeq\frac1n\sum_{v\in V}P^\mu_v$.
More explicitly, $P^\mu$ is the transition matrix of the Markov chain that moves from the current configuration $\sigma\in\Omega$ to a new configuration $\sigma'\in\Omega$ as follows:
\begin{enumerate}[label=\textup{(\arabic*)}]
  \item pick a vertex $v\in V$ uniformly at random;
  \item sample $\sigma'_v\sim\mu(\sigma_v=\cdot\mid\sigma_{V\setminus\{v\}})$, and set $\sigma'_u\defeq\sigma_u$ for all $u\in V\setminus\{v\}$.
\end{enumerate}

As an operator on $L^2(\mu)$, $P^\mu_v$ takes the conditional expectation given $\sigma_{V\setminus\{v\}}$,
and is therefore the orthogonal projection onto functions that do not depend on $\sigma_v$.
Consequently, $P^\mu_v$ and $P^\mu$ are self-adjoint and positive semidefinite on $L^2(\mu)$,
so $P^\mu$ is reversible with respect to $\mu$ and has no negative eigenvalues.
The complementary projection $\Id-P^\mu_v$ subtracts this conditional expectation, so every $f\in L^2(\mu)$ satisfies
\begin{equation}\label{eq:heat-bath-form}
  \ip[\mu]f{\tuple{\Id-P^\mu_v}f}=\norm[\mu]{\tuple{\Id-P^\mu_v}f}^2
  =\E[\sigma\sim\mu]{
    \Var[\tau\sim\mu]{
      f\tuple{\tau}\mid
      \tau_{V\setminus\{v\}}=\sigma_{V\setminus\{v\}}
    }
  }.
\end{equation}

\subsection{Proper colorings}\label{sec:prelim-colorings}

Let $G=(V,E)$ be a finite simple graph with $n\ge 1$ vertices and maximum degree at most $\Delta$.
Write $N(v)$ for the set of neighbors of a vertex $v$, $\deg(v)\defeq|N(v)|$, $u\sim v$ when $uv\in E$, and $u\not\sim v$ when $u\neq v$ and $uv\notin E$.
A \emph{proper coloring} of $G$, from now on simply called a \emph{coloring},
is a map $\sigma\colon V\to[q]$ such that $\sigma_u\ne\sigma_v$ for every edge $uv$.
Let $\Omega$ be the set of colorings of $G$, and let $\mu$ be the uniform distribution on $\Omega$.
Throughout the rest of this paper, we assume that $q\ge\Delta+3$.

For a coloring $\sigma$ and a vertex $v$, let $A_v(\sigma)$ be the set of colors available to $v$ in $\sigma$, i.e., the colors not used by any neighbor of $v$.
Note that $A_v(\sigma)$ contains $\sigma_v$, depends only on $\sigma_{V\setminus\{v\}}$, and has size at least $q-\Delta$.
For $c\in[q]$, write $A^{-c}_v(\sigma)\defeq A_v(\sigma)\setminus\{c\}$.
For a color $c\in A_v(\sigma)$, let $\sigma^{v\gets c}$ be the coloring obtained from $\sigma$ by recoloring $v$ with $c$.

Under $\mu$, conditionally on $\sigma_{V\setminus\{v\}}$, the color $\sigma_v$ is uniform on $A_v(\sigma)$, where each available color has probability
\begin{equation*}
  p_v\tuple{\sigma}\defeq\frac1{|A_v\tuple{\sigma}|}.
\end{equation*}
Like $A_v(\sigma)$, the probability $p_v(\sigma)$ depends only on $\sigma_{V\setminus\{v\}}$.
Writing $P_v\defeq P^\mu_v$ for the transition matrix of the heat-bath update at $v$,
we have $(P_vf)(\sigma)=p_v(\sigma)\sum_{c\in A_v(\sigma)}f(\sigma^{v\gets c})$.
The \emph{Glauber dynamics} for colorings is $P\defeq P^\mu=\frac1n\sum_{v\in V}P_v$.
It moves from the current coloring $\sigma\in\Omega$ as follows:
\begin{enumerate}[label=\textup{(\arabic*)}]
  \item pick a vertex $v\in V$ uniformly at random;
  \item pick a color $c\in A_v(\sigma)$ uniformly at random, and move to $\sigma^{v\gets c}$.
\end{enumerate}

\begin{fact}[\cite{Jer95Very}]\label{fact:irreducible}
  If $q\ge\Delta+2$, then $P$ is irreducible.
\end{fact}

Let $Q_v\defeq\Id-P_v$ and $L\defeq\sum_vQ_v=n(\Id-P)$.
By \pref{sec:prelim-gap}, every $Q_v$ is an orthogonal projection, and $L$ is self-adjoint and positive semidefinite on $L^2(\mu)$.
The spectral gap of $P$ is certified by the following \emph{integrated Bochner inequality}.

\begin{lemma}\label{lem:spectral}
  For every $\kappa>0$, $\gap(P)\ge\kappa/n$ if and only if for every $f\in L^2(\mu)$,
  \begin{equation*}
    \norm{Lf}^2\ge\kappa\ip f{Lf}.
  \end{equation*}
\end{lemma}

\begin{proof}
  Let $(e_j)_{j\ge1}$ be an orthonormal basis of $L^2(\mu)$ with $Pe_j=\lambda_j(P)e_j$, and set $z_j\defeq n(1-\lambda_j(P))$,
  so that $Le_j=z_je_j$ and $0=z_1\le z_2\le\cdots$.
  For $f=\sum_jb_je_j$, we have $\norm{Lf}^2-\kappa\ip f{Lf}=\sum_jz_j(z_j-\kappa)b_j^2$,
  so the integrated Bochner inequality holds for every $f\in L^2(\mu)$ if and only if each $z_j$ is either $0$ or at least $\kappa$.
  By \pref{fact:irreducible}, the eigenvalue $1$ of $P$ is simple, so $z_j>0$ for every $j\ge2$, and the condition is equivalent to $z_2=n\gap(P)\ge\kappa$.
\end{proof}

By \pref{eq:heat-bath-form}, the Dirichlet form of $L$ decomposes as $\ip f{Lf}=\sum_v\norm{Q_vf}^2$.
Since $\sigma_v$ is uniform on $A_v(\sigma)$ conditionally on the colors off $v$,
and the expectation of the squared difference of two independent copies of a random variable is twice its variance, we have
\begin{equation}\label{eq:first-diff}
  \norm{Q_vf}^2=\frac 12\sum_{\sigma\in\Omega}\mu\tuple{\sigma}\,p_v\tuple{\sigma}\sum_{c\in A_v\tuple{\sigma}}\tuple{f\tuple{\sigma}-f\tuple{\sigma^{v\gets c}}}^2.
\end{equation}

\subsection{The hard-core model}\label{sec:prelim-hardcore}

Let $\+I$ be the set of independent sets of $G$, including the empty set.
In this subsection, fix activities $\lambda=(\lambda_v)_{v\in V}$ with $\lambda_v>0$.
We suppress the dependence on $\lambda$ in the notation.
The \emph{hard-core model} on $G$ is the probability distribution $\nu$ on $\+I$ given by
\begin{equation}\label{eq:hardcore}
  \nu\tuple{I}\defeq\frac1{Z}\prod_{v\in I}\lambda_v,\qquad
  Z\defeq\sum_{I'\in\+I}\prod_{v\in I'}\lambda_v.
\end{equation}

For an independent set $I$, a vertex $v$ is \emph{occupied} if $v\in I$, and \emph{unblocked} if it has no occupied neighbor;
let $V_{\mathrm{ubl}}(I)\defeq\{v\in V:N(v)\cap I=\varnothing\}$.
We call the removal of an occupied vertex a \emph{death} and the addition of an unoccupied, unblocked vertex a \emph{birth}.
Let $p_v\defeq\lambda_v/(1+\lambda_v)$.
Given the occupation off $v$, the vertex $v$ must be unoccupied if $v\notin V_{\mathrm{ubl}}(I)$,
whereas if $v\in V_{\mathrm{ubl}}(I)$ it is occupied with conditional probability $\nu(I\cup\{v\})/(\nu(I\cup\{v\})+\nu(I\setminus\{v\}))=p_v$ by \pref{eq:hardcore}.
Thus the Glauber dynamics for $\nu$, with transition matrix $P^\nu$, moves from the current independent set $I\in\+I$ as follows:
\begin{enumerate}[label=\textup{(\arabic*)}]
  \item pick a vertex $v\in V$ uniformly at random;
  \item if $v\in V_{\mathrm{ubl}}(I)$, move to $I\cup\{v\}$ with probability $p_v$ and to $I\setminus\{v\}$ with probability $1-p_v$;
    otherwise, stay at $I$.
\end{enumerate}
Set $H_v\defeq\Id-P^\nu_v$, $H\defeq\sum_{v\in V}H_v=n(\Id-P^\nu)$, and $\gap(H)\defeq n\gap(P^\nu)$.
By \pref{sec:prelim-gap}, every $H_v$ is an orthogonal projection on $L^2(\nu)$, and $H$ is self-adjoint and positive semidefinite on $L^2(\nu)$.
Explicitly, for every $h\in L^2(\nu)$,
\begin{equation}\label{eq:hc-generator}
  \tuple{H h}\tuple{I}=\sum_{v\in V_{\mathrm{ubl}}\tuple{I}}\tuple{h\tuple{I}-p_vh\tuple{I\cup\{v\}}-\tuple{1-p_v}h\tuple{I\setminus\{v\}}}.
\end{equation}
By \pref{eq:heat-bath-form} and the conditional distribution at an unblocked vertex, the Dirichlet form of $H_v$ can be written as
\begin{align}
  \ip[\nu]h{H_vh}
  &=\E[I\sim\nu]{\one\{v\in I\}\tuple{1-p_v}\tuple{h\tuple{I}-h\tuple{I\setminus\{v\}}}^2}\label{eq:hc-death}\\
  &=\E[I\sim\nu]{\one\{v\in V_{\mathrm{ubl}}\tuple{I}\setminus I\}\,p_v\tuple{h\tuple{I}-h\tuple{I\cup\{v\}}}^2}\label{eq:hc-birth}\\
  &=\E[I\sim\nu]{\one\{v\in V_{\mathrm{ubl}}\tuple{I}\}\,p_v\tuple{1-p_v}\tuple{h\tuple{I\cup\{v\}}-h\tuple{I\setminus\{v\}}}^2}.\label{eq:hc-unblock}
\end{align}

Let $\lambda_{\mathrm{crit}}(\Delta)\defeq\frac{(\Delta-1)^{\Delta-1}}{(\Delta-2)^\Delta}$ be the uniqueness threshold of the hard-core model on the infinite $\Delta$-regular tree;
note that $\lambda_{\mathrm{crit}}(\Delta)\sim \ee/\Delta$ as $\Delta\to\infty$.
Chen, Feng, Yin, and Zhang~\cite{CFYZ21Rapid} proved that, uniformly below this threshold,
$\gap(H)$ admits a constant lower bound independent of both the graph and its maximum degree $\Delta$.\footnote{More precisely,
  their result is stated for a uniform activity on graphs of maximum degree exactly~$\Delta$.
  However, the form stated in \pref{thm:hardcore}, with vertex-dependent activities and maximum degree at most~$\Delta$, follows easily from their argument.}
  This uniformity in $\Delta$ is essential for our argument;
  by contrast, the bounds in~\cite{JPV22Spectral,CLV21Optimal} degrade with $\Delta$.
\begin{theorem}[\cite{CFYZ21Rapid}]\label{thm:hardcore}
  For every $\delta\in(0,1)$ there is $\theta_*(\delta)\in(0,1]$ such that, for every $\Delta\ge3$, 
  every $n$-vertex graph of maximum degree at most $\Delta$,
  and all activities $0<\lambda_v\le(1-\delta)\lambda_{\mathrm{crit}}(\Delta)$, 
  we have $\gap(P^{\nu})\ge\theta_*(\delta)/n$, and hence $\gap(H)\ge\theta_*(\delta)$.
\end{theorem}

\section{Single-color decomposition}\label{sec:decomposition}

In this section, we introduce the single-color decomposition of $L$, state two comparison lemmas for the resulting single-color operators,
and deduce \pref{thm:main} from them.

\subsection{Single-color operators}\label{sec:stars}

Fix a vertex $v\in V$ and a color $c\in[q]$.
Let $P_{v,c}$ be the transition matrix on $\Omega$ that keeps only the moves of the heat-bath update $P_v$ whose old or new color is $c$, i.e.,
for $\sigma\in\Omega$ and $c'\in A^{-\sigma_v}_v(\sigma)$,
\begin{equation}\label{eq:star-transition}
  P_{v,c}\tuple{\sigma,\sigma^{v\gets c'}}\defeq p_v\tuple{\sigma}\,\one\{c\in\{\sigma_v,c'\}\},
\end{equation}
the diagonal entry $P_{v,c}(\sigma,\sigma)$ is the remaining probability, and all other entries of $P_{v,c}$ are zero.
Equivalently, $P_{v,c}$ moves from the current coloring $\sigma\in\Omega$ as follows:
\begin{enumerate}[label=\textup{(\arabic*)}]
  \item pick a color $c'\in A_v(\sigma)$ uniformly at random;
  \item if $\sigma_v=c$ or $c'=c$, move to $\sigma^{v\gets c'}$;
    otherwise, stay at $\sigma$.
\end{enumerate}
Recoloring $v$ does not change $A_v(\sigma)$, and the condition $c\in\{\sigma_v,c'\}$ is symmetric in the old and new colors,
so a move and its reverse have the same transition probability.
Since $\mu$ is uniform, $P_{v,c}$ is therefore reversible with respect to $\mu$,
and by \pref{sec:prelim-gap} it is self-adjoint on $L^2(\mu)$ with all eigenvalues in $[-1,1]$.
Hence $Q_{v,c}\defeq\Id-P_{v,c}$ is self-adjoint and positive semidefinite on $L^2(\mu)$.
Explicitly,
\begin{equation}\label{eq:star-def}
  \tuple{Q_{v,c}f}\tuple{\sigma}
  =\begin{cases}
    \displaystyle\tuple{Q_vf}\tuple{\sigma}=p_v\tuple{\sigma}\sum_{c'\in A_v\tuple{\sigma}}\tuple{f\tuple{\sigma}-f\tuple{\sigma^{v\gets c'}}},&\sigma_v=c,\\[2.2ex]
    \displaystyle p_v\tuple{\sigma}\tuple{f\tuple{\sigma}-f\tuple{\sigma^{v\gets c}}},&\sigma_v\ne c\text{ and }c\in A_v\tuple{\sigma},\\[2.2ex]
    0,&c\notin A_v\tuple{\sigma}.
  \end{cases}
\end{equation}

With $\sigma_{V\setminus\{v\}}$ fixed, we may identify each available color $c'\in A_v(\sigma)$ with the coloring $\sigma^{v\gets c'}$.
The moves of $P_v$ then connect every two available colors, with transition probability $p_v(\sigma)$ in each direction.
For $c\in A_v(\sigma)$, the moves retained by $P_{v,c}$ are those incident to $c$;
they form a star centered at $c$, and we therefore call $Q_{v,c}$ a \emph{star operator}.
Every nontrivial recoloring at $v$ belongs to exactly two of these stars, namely those centered at its old and its new color,
so the star operators at $v$ sum to $2Q_v$;
see \pref{fig:stars} and \pref{lem:star} below.

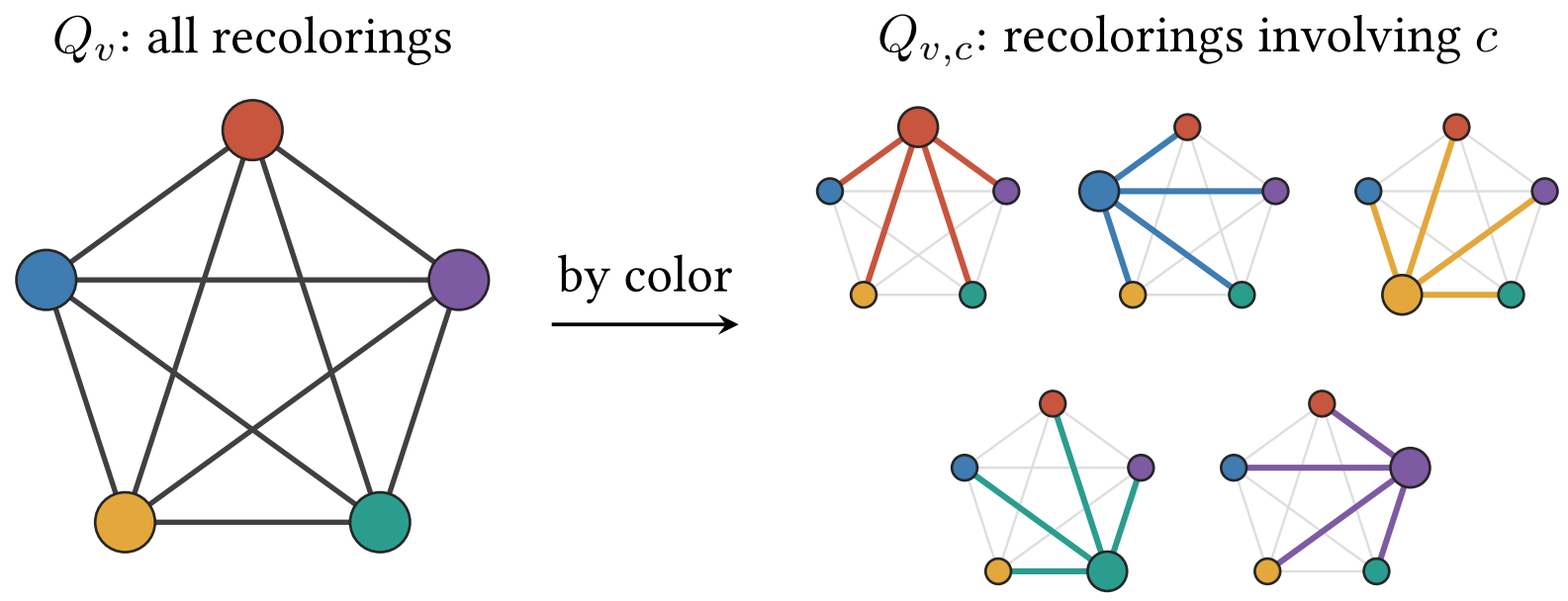
\begin{figure}[H]
  \centering
  \definecolor{StarRed}{HTML}{C8553D}
  \definecolor{StarBlue}{HTML}{3E7CB1}
  \definecolor{StarGold}{HTML}{E3A73B}
  \definecolor{StarTeal}{HTML}{2A9D8F}
  \definecolor{StarPurple}{HTML}{7E5AA2}
  \begin{tikzpicture}[
      font=\small,
      kedge/.style={draw=black!75, line width=1.0pt},
      kfaint/.style={draw=black!13, line width=0.45pt},
      kdot/.style={draw=black!85, line width=0.5pt},
    ]
    \def\StarGraph#1#2#3#4#5{\begin{scope}[shift={#1}]
        \foreach \i in {1,...,4}{\pgfmathtruncatemacro{\inext}{\i+1}\foreach \j in {\inext,...,5}{\pgfmathtruncatemacro{\instar}{(\i==#4)||(\j==#4)}\ifnum\instar=0
              \ifnum#4=0
                \draw[kedge] (18+72*\i:#2) -- (18+72*\j:#2);
              \else
                \draw[kfaint] (18+72*\i:#2) -- (18+72*\j:#2);
              \fi
            \fi
        }}\ifnum#4>0
          \foreach \j in {1,...,5}{\ifnum\j=#4\else
              \draw[draw=#5, line width=1.15pt] (18+72*#4:#2) -- (18+72*\j:#2);
            \fi
        }\fi
        \foreach \i/\c in {1/StarRed,2/StarBlue,3/StarGold,4/StarTeal,5/StarPurple}{\ifnum\i=#4
            \filldraw[kdot, fill=\c] (18+72*\i:#2) circle[radius=1.55*#3];
          \else
            \filldraw[kdot, fill=\c] (18+72*\i:#2) circle[radius=#3];
          \fi
    }\end{scope}}
    \StarGraph{(0,0)}{1.45}{0.2}{0}{black}
    \node at (0,2.05) {$Q_v$: all recolorings};
    \draw[-stealth, line width=0.7pt] (2.0,0.15) -- node[above] {by color} (3.25,0.15);
    \node at (6.25,2.05) {$Q_{v,c}$: recolorings involving $c$};
    \StarGraph{(4.45,0.85)}{0.62}{0.085}{1}{StarRed}
    \StarGraph{(6.25,0.85)}{0.62}{0.085}{2}{StarBlue}
    \StarGraph{(8.05,0.85)}{0.62}{0.085}{3}{StarGold}
    \StarGraph{(5.35,-1.0)}{0.62}{0.085}{4}{StarTeal}
    \StarGraph{(7.15,-1.0)}{0.62}{0.085}{5}{StarPurple}
  \end{tikzpicture}
  \caption{The star operators at a vertex $v$, with $\sigma_{V\setminus\{v\}}$ fixed and $|A_v(\sigma)|=5$.
    On the left, the moves of $P_v$ form a complete graph on the five available colors,
    and each edge is traversed with probability $p_v(\sigma)=1/5$ in either direction.
    On the right, for each available color $c$, the moves retained by $P_{v,c}$ form the star of edges incident to $c$,
    drawn in the color $c$ with an enlarged center;
    the remaining edges are drawn faintly.}
  \label{fig:stars}
\end{figure}

Just as the operators $Q_v$ sum to $L$, summing the star operators over $v$ gives the \emph{single-color operator} $L_c\defeq\sum_vQ_{v,c}$,
which retains exactly the moves of $L$ whose old or new color is $c$.
As a sum of the operators $Q_{v,c}$, it is self-adjoint and positive semidefinite on $L^2(\mu)$.
The following lemma relates these operators to $Q_v$ and $L$.

\begin{lemma}\label{lem:star}
  For every vertex $v\in V$, we have $\sum_{c\in[q]}Q_{v,c}=2Q_v$ and $\sum_{c\in[q]}L_c=2L$.
\end{lemma}

\begin{proof}
  Fix $\sigma\in\Omega$ and $c'\in A^{-\sigma_v}_v(\sigma)$.
  Since $c'\ne\sigma_v$, exactly two colors $c\in[q]$ satisfy $c\in\{\sigma_v,c'\}$, namely the old and the new color of the move to $\sigma^{v\gets c'}$.
  Hence \pref{eq:star-transition} gives
  \begin{equation*}
    \sum_{c\in[q]}P_{v,c}\tuple{\sigma,\sigma^{v\gets c'}}=2p_v\tuple{\sigma}=2P_v\tuple{\sigma,\sigma^{v\gets c'}}.
  \end{equation*}
  Thus the matrices $\sum_cQ_{v,c}$ and $2Q_v$ have the same off-diagonal entries.
  Since both have zero row sums, their diagonal entries agree as well, which proves the first identity.
  Summing it over $v$ gives the second.
\end{proof}

\subsection{Two comparisons: Proof of \texorpdfstring{\pref{thm:main}}{Theorem \getrefnumber{thm:main}}}\label{sec:comparisons}

With the single-color operators $L_c$ in place, we can now make our two-step strategy precise.
Our goal is to establish the two inequalities
\begin{equation*}
  \norm{Lf}^2\gtrsim\sum_{c\in[q]}\norm{L_cf}^2\gtrsim_\gamma\ip f{Lf}.
\end{equation*}

The first step is a local comparison that bounds $\sum_c\norm{L_cf}^2$ from above in terms of $\norm{Lf}^2$.
We prove it in \pref{sec:bochner} by expanding both sides into contributions from individual vertices and pairs of vertices,
and comparing these contributions term by term.

\begin{lemma}\label{lem:bochner-comparison}
  Suppose that $\Delta\ge1$ and $q\ge\Delta+3$.
  Then, for every $f\in L^2(\mu)$,
  \begin{equation}\label{eq:bochner-comparison}
    \sum_{c\in[q]}\norm{L_cf}^2\le\tuple{1+\frac2{q-\Delta}}\norm{Lf}^2
    +\frac{50\Delta}{9\tuple{q-\Delta}^2}\ip f{Lf}.
  \end{equation}
\end{lemma}

The second step is a global comparison that bounds $\sum_c\norm{L_cf}^2$ from below in terms of $\ip f{Lf}$.
We prove it in \pref{sec:hidden} and~\pref{sec:resolvent} by exploiting the spectral gaps of the hard-core models hidden in the color classes.
The term $1/\ee$ in $(1+1/\ee)\Delta$ comes precisely from the uniqueness threshold of the hard-core model.

\begin{lemma}\label{lem:hardcore-comparison}
  For every $\gamma>0$, there is a constant $\theta_\gamma\in(0,1]$ such that if $q\ge(1+1/\ee+\gamma)\Delta$ and $\Delta\ge\max\{3,2/\gamma\}$,
  then for every $f\in L^2(\mu)$,
  \begin{equation}\label{eq:hardcore-comparison}
    \sum_{c\in[q]}\norm{L_cf}^2\ge\frac{\theta_\gamma}{24}\ip f{Lf}
    -O_\gamma\tuple{\frac1\Delta}\tuple{\norm{Lf}^2+\ip f{Lf}}.
  \end{equation}
\end{lemma}

Combining these two comparisons immediately yields the main theorem.

\begin{proof}[Proof of \pref{thm:main}]
  Fix $\gamma>0$, and let $q\ge(1+1/\ee+\gamma)\Delta$, where $\Delta$ is sufficiently large in terms of $\gamma$.
  Then $q\ge\Delta+3$ and $q-\Delta\ge\Delta/\ee$, so \pref{lem:bochner-comparison} gives
  \begin{equation*}
    \sum_{c\in[q]}\norm{L_cf}^2\le\tuple{1+\frac{2\ee}{\Delta}}\norm{Lf}^2
    +\frac{50\ee^2}{9\Delta}\ip f{Lf}.
  \end{equation*}
  Combining this with the lower bound in \pref{lem:hardcore-comparison} and rearranging, we obtain
  \begin{equation*}
    \norm{Lf}^2\ge\frac{\theta_\gamma}{24}\ip f{Lf}
    -\frac{C_\gamma}\Delta\tuple{\norm{Lf}^2+\ip f{Lf}},
  \end{equation*}
  where $C_\gamma>0$ depends only on $\gamma$.
  If $\Delta\ge48C_\gamma/\theta_\gamma$, then $C_\gamma/\Delta\le\theta_\gamma/48\le1$, and rearranging yields
  \begin{equation*}
    \frac{\theta_\gamma}{48}\ip f{Lf}\le\tuple{1+\frac{C_\gamma}\Delta}\norm{Lf}^2\le2\norm{Lf}^2.
  \end{equation*}
  Thus $\norm{Lf}^2\ge\kappa_\gamma\ip f{Lf}$ for every $f\in L^2(\mu)$, where $\kappa_\gamma\defeq\theta_\gamma/96$.
  By \pref{fact:irreducible}, $P$ is irreducible, so \pref{lem:spectral} gives $\gap(P)\ge\kappa_\gamma/n$.
  Finally, $P$ is positive semidefinite and $\mu$ is uniform on $\Omega$ with $|\Omega|\le q^n$, so the mixing-time bound follows from \pref{lem:mixing}.
\end{proof}

\section{The local comparison}\label{sec:bochner}

In this section, we prove \pref{lem:bochner-comparison} by expanding $\norm{Lf}^2$ and $\sum_c\norm{L_cf}^2$ into terms indexed by vertices and by pairs of vertices,
and comparing the two expansions term by term.
Since $L=\sum_vQ_v$ and $L_c=\sum_vQ_{v,c}$, we have
\begin{align}
  \norm{Lf}^2&=\sum_{v\in V}\norm{Q_vf}^2+\sum_{u\ne v}\ip{Q_uf}{Q_vf},\label{eq:expand-L}\\
  \sum_{c\in[q]}\norm{L_cf}^2&=\sum_{v\in V}\sum_{c\in[q]}\norm{Q_{v,c}f}^2
  +\sum_{u\ne v}\sum_{c\in[q]}\ip{Q_{u,c}f}{Q_{v,c}f}.\label{eq:expand-R}
\end{align}
The following three lemmas compare the single-vertex terms, the cross terms of non-adjacent pairs,
and the cross terms of adjacent pairs in \pref{eq:expand-R} with their counterparts in \pref{eq:expand-L}.
We first derive \pref{lem:bochner-comparison} from them
and then prove them in \pref{sec:single-vertex}, \pref{sec:nonadjacent}, and~\pref{sec:adjacent}, respectively.

\begin{lemma}\label{lem:single-vertex}
  For every vertex $v\in V$ and every $f\in L^2(\mu)$,
  \begin{equation*}
    \norm{Q_vf}^2\le\sum_{c\in[q]}\norm{Q_{v,c}f}^2
    \le\tuple{1+\frac2{q-\Delta}}\norm{Q_vf}^2.
  \end{equation*}
\end{lemma}

\begin{lemma}\label{lem:nonedge}
  For distinct non-adjacent $u,v\in V$ and every $f\in L^2(\mu)$,
  \begin{equation*}
    0\le\sum_{c\in[q]}\ip{Q_{u,c}f}{Q_{v,c}f}\le\tuple{1+\frac2{q-\Delta}}\ip{Q_uf}{Q_vf}.
  \end{equation*}
\end{lemma}

\begin{lemma}\label{lem:edge}
  For every edge $uv\in E$ and every $f\in L^2(\mu)$,
  \begin{equation*}
    \begin{aligned}
      \sum_{c\in[q]}\ip{Q_{u,c}f}{Q_{v,c}f}&\ge-\frac1{q-\Delta}\tuple{\norm{Q_uf}^2+\norm{Q_vf}^2},\\
      \sum_{c\in[q]}\ip{Q_{u,c}f}{Q_{v,c}f}&\le\ip{Q_uf}{Q_vf}+\frac4{9\tuple{q-\Delta}^2}\tuple{\norm{Q_uf}^2+\norm{Q_vf}^2}.
    \end{aligned}
  \end{equation*}
\end{lemma}

\begin{proof}[Proof of \pref{lem:bochner-comparison}]
  Since every vertex has at most $\Delta$ neighbors and $\sum_v\norm{Q_vf}^2=\ip f{Lf}$, we have
  \begin{equation}\label{eq:edge-count}
    \sum_{u\sim v}\tuple{\norm{Q_uf}^2+\norm{Q_vf}^2}\le2\Delta\ip f{Lf}.
  \end{equation}
  We subtract \pref{eq:expand-L} from \pref{eq:expand-R} and apply \pref{lem:single-vertex} to the single-vertex terms,
  \pref{lem:nonedge} to the cross terms of non-adjacent pairs, and the upper bound in \pref{lem:edge} to those of adjacent pairs.
  Together with \pref{eq:edge-count}, this gives
  \begin{equation*}
    \sum_{c\in[q]}\norm{L_cf}^2-\norm{Lf}^2
    \le\frac2{q-\Delta}\tuple{\ip f{Lf}+\sum_{u\not\sim v}\ip{Q_uf}{Q_vf}}+\frac{8\Delta}{9\tuple{q-\Delta}^2}\ip f{Lf}.
  \end{equation*}
  To bound the bracketed term, we combine the two bounds of \pref{lem:edge}, which give, for every edge $uv$,
  \begingroup
    \medmuskip=2mu
    \thickmuskip=3mu
    \begin{equation*}
      \ip{Q_uf}{Q_vf}\ge-\tuple{\frac1{q-\Delta}+\frac4{9\tuple{q-\Delta}^2}}\tuple{\norm{Q_uf}^2+\norm{Q_vf}^2}
      \ge-\frac7{6\tuple{q-\Delta}}\tuple{\norm{Q_uf}^2+\norm{Q_vf}^2},
    \end{equation*}
  \endgroup
  where the last step uses $q-\Delta\ge3$.
  Together with \pref{eq:expand-L} and \pref{eq:edge-count}, this gives
  \begin{equation*}
    \ip f{Lf}+\sum_{u\not\sim v}\ip{Q_uf}{Q_vf}
    =\norm{Lf}^2-\sum_{u\sim v}\ip{Q_uf}{Q_vf}
    \le\norm{Lf}^2+\frac{7\Delta}{3\tuple{q-\Delta}}\ip f{Lf}.
  \end{equation*}
  Substituting this bound into the estimate for $\sum_c\norm{L_cf}^2-\norm{Lf}^2$ above yields
  \begin{equation*}
    \sum_{c\in[q]}\norm{L_cf}^2-\norm{Lf}^2\le\frac2{q-\Delta}\norm{Lf}^2+\frac{50\Delta}{9\tuple{q-\Delta}^2}\ip f{Lf}.\qedhere
  \end{equation*}
\end{proof}

\subsection{Single-vertex terms}\label{sec:single-vertex}

\begin{proof}[Proof of \pref{lem:single-vertex}]
  Fix $\sigma\in\Omega$.
  By \pref{eq:star-def}, $(Q_{v,c}f)(\sigma)$ equals $(Q_vf)(\sigma)$ for $c=\sigma_v$,
  equals $p_v(\sigma)(f(\sigma)-f(\sigma^{v\gets c}))$ for every other $c\in A_v(\sigma)$, and vanishes for $c\notin A_v(\sigma)$, so
  \begin{equation*}
    \sum_{c\in[q]}\tuple{Q_{v,c}f}\tuple{\sigma}^2
    =\tuple{Q_vf}\tuple{\sigma}^2+p_v\tuple{\sigma}^2\sum_{c\in A_v\tuple{\sigma}}\tuple{f\tuple{\sigma}-f\tuple{\sigma^{v\gets c}}}^2.
  \end{equation*}
  As in the derivation of \pref{eq:first-diff}, conditionally on $\sigma_{V\setminus\{v\}}$,
  the expectation of $\frac12p_v(\sigma)\sum_{c\in A_v(\sigma)}(f(\sigma)-f(\sigma^{v\gets c}))^2$ equals that of $(Q_vf)(\sigma)^2$,
  namely the conditional variance of $f$.
  Since $p_v(\sigma)$ depends only on $\sigma_{V\setminus\{v\}}$, taking the expectation over $\sigma\sim\mu$ gives
  \begin{equation*}
    \sum_{c\in[q]}\norm{Q_{v,c}f}^2=\sum_{\sigma\in\Omega}\mu\tuple{\sigma}\tuple{1+2p_v\tuple{\sigma}}\tuple{Q_vf}\tuple{\sigma}^2.
  \end{equation*}
  Since $0<p_v(\sigma)\le1/(q-\Delta)$, both bounds follow.
\end{proof}

\subsection{Non-adjacent pairs}\label{sec:nonadjacent}

Let $u$ and $v$ be distinct non-adjacent vertices.
Recoloring either vertex does not change the colors available at the other.
Hence, for $c\in A_u(\sigma)$ and $c'\in A_v(\sigma)$, recoloring $u$ and $v$ in either order yields the same coloring,
which we denote by $\sigma^{u\gets c,\,v\gets c'}$, and conditionally on $\sigma_{V\setminus\{u,v\}}$,
the pair $(\sigma_u,\sigma_v)$ is uniform on $A_u(\sigma)\times A_v(\sigma)$.
Every update operator at $u$, namely $P_u$ or some $P_{u,c}$, recolors only $u$,
and by \pref{eq:star-transition} its transition probabilities depend only on the old and new colors at $u$ and on $A_u(\sigma)$,
which is unaffected by recoloring $v$;
the same holds at $v$.
Therefore, applying an update at $u$ followed by one at $v$, or vice versa, moves from $\sigma$ to $\sigma^{u\gets c,\,v\gets c'}$ with the same probability,
namely the product of the two individual transition probabilities.
Thus every update operator at $u$ commutes with every update operator at $v$, and so do the corresponding operators $Q=\Id-P$.

\begin{proof}[Proof of \pref{lem:nonedge}]
  Fix a color $c$.
  The operators $Q_{u,c}$ and $Q_{v,c}$ commute and are self-adjoint and positive semidefinite, so their product is positive semidefinite as well.
  Hence $\ip{Q_{u,c}f}{Q_{v,c}f}=\ip f{Q_{u,c}Q_{v,c}f}\ge0$, and summing over $c$ proves the lower bound.

  For the upper bound, note that $P_{v,c}$ recolors only $v$, so it fixes every function that does not depend on $\sigma_v$, i.e.,
  every function in the range of $P_v$.
  Thus $Q_{v,c}P_v=0$, and taking adjoints gives $P_vQ_{v,c}=0$.
  Since $Q_v=\Id-P_v$, it follows that $Q_{v,c}=Q_{v,c}Q_v=Q_vQ_{v,c}$, and similarly $Q_{u,c}=Q_{u,c}Q_u=Q_uQ_{u,c}$.
  Together with self-adjointness and commutativity, this gives
  \begin{equation*}
    \ip{Q_{u,c}f}{Q_{v,c}f}=\ip{Q_uQ_{u,c}f}{Q_vQ_{v,c}f}=\ip{Q_{u,c}Q_vf}{Q_{v,c}Q_uf}.
  \end{equation*}
  Moreover, since $Q_u$ and $Q_v$ are commuting orthogonal projections,
  \begin{equation}\label{eq:nonedge-positive}
    \ip{Q_uf}{Q_vf}=\ip f{Q_uQ_vf}=\ip f{Q_vQ_u^2Q_vf}=\norm{Q_uQ_vf}^2.
  \end{equation}
  By the Cauchy--Schwarz inequality and \pref{lem:single-vertex} applied at $u$ to $Q_vf$ and at $v$ to $Q_uf$,
  \begin{align*}
    \sum_{c\in[q]}\ip{Q_{u,c}f}{Q_{v,c}f}
    &\le\frac12\sum_{c\in[q]}\norm{Q_{u,c}Q_vf}^2
    +\frac12\sum_{c\in[q]}\norm{Q_{v,c}Q_uf}^2\\
    &\le\tuple{1+\frac2{q-\Delta}}\norm{Q_uQ_vf}^2
    =\tuple{1+\frac2{q-\Delta}}\ip{Q_uf}{Q_vf}.\qedhere
  \end{align*}
\end{proof}

\subsection{Adjacent pairs}\label{sec:adjacent}

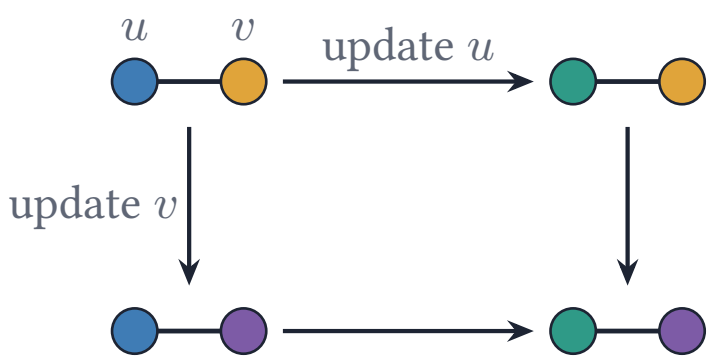
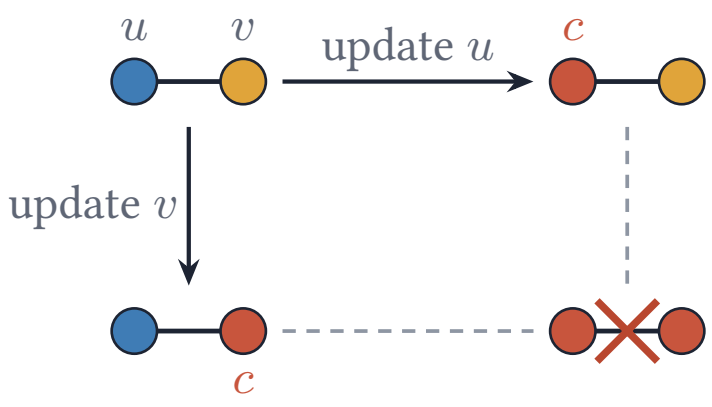
\begin{figure}[htbp]
  \centering
  \begin{subfigure}[t]{0.48\linewidth}
    \centering
    \begin{tikzpicture}[x=1cm,y=1cm]
      \path[use as bounding box] (-1.5,-2.25) rectangle (3.65,0.6);
      \node[flab] at (-0.36,0.34) {$u$};
      \node[flab] at (0.36,0.34) {$v$};
      \pic at (0,0)      {pair={gblue}{ggold}};
      \pic at (2.9,0)    {pair={gteal}{ggold}};
      \pic at (0,-1.65)  {pair={gblue}{gpurple}};
      \pic at (2.9,-1.65){pair={gteal}{gpurple}};
      \draw[move] (0.62,0) -- node[above,flab]{update $u$} (2.28,0);
      \draw[move] (0,-0.3) -- node[left,flab]{update $v$} (0,-1.35);
      \draw[move] (0.62,-1.65) -- (2.28,-1.65);
      \draw[move] (2.9,-0.3) -- (2.9,-1.35);
    \end{tikzpicture}
    \caption{Different new colors.}
    \label{fig:rectangle-intro-different}
  \end{subfigure}\hfill\begin{subfigure}[t]{0.48\linewidth}
    \centering
    \begin{tikzpicture}[x=1cm,y=1cm]
      \path[use as bounding box] (-1.5,-2.25) rectangle (3.65,0.6);
      \node[flab] at (-0.36,0.34) {$u$};
      \node[flab] at (0.36,0.34) {$v$};
      \pic at (0,0)      {pair={gblue}{ggold}};
      \pic at (2.9,0)    {pair={gx}{ggold}};
      \pic at (0,-1.65)  {pair={gblue}{gx}};
      \pic at (2.9,-1.65){pair={gx}{gx}};
      \node[flab,text=gx] at (2.54,0.34) {$c$};
      \node[flab,text=gx] at (0.36,-1.99) {$c$};
      \draw[move] (0.62,0) -- node[above,flab]{update $u$} (2.28,0);
      \draw[move] (0,-0.3) -- node[left,flab]{update $v$} (0,-1.35);
      \draw[nomove] (0.62,-1.65) -- (2.28,-1.65);
      \draw[nomove] (2.9,-0.3) -- (2.9,-1.35);
      \draw[cterra,line width=1.4pt]
        (2.7,-1.85) -- (3.1,-1.45) (2.7,-1.45) -- (3.1,-1.85);
    \end{tikzpicture}
    \caption{A common new color $c$.}
    \label{fig:rectangle-intro-same}
  \end{subfigure}
  \caption{Two individually legal recolorings at the endpoints of an edge $uv$, with all other colors fixed;
    horizontal arrows recolor $u$, and vertical arrows recolor $v$.
    (a) If the new colors differ, performing both recolorings gives a proper coloring, and the four colorings form a legal rectangle.
    (b) If both new colors equal $c$, the fourth coloring is not proper, and the dashed recolorings are not available;
    the corresponding cross terms are those of the single-color operators $L_c$.}
  \label{fig:rectangle-intro}
\end{figure}

Fix an edge $uv\in E$.
We split $\ip{Q_uf}{Q_vf}$ according to whether the two proposed new colors agree, and write $\omega_{uv}(\sigma)\defeq\mu(\sigma)\,p_u(\sigma)\,p_v(\sigma)$.
Since $(Q_uf)(\sigma)=p_u(\sigma)\sum_{c_u\in A_u(\sigma)}(f(\sigma)-f(\sigma^{u\gets c_u}))$, and similarly at $v$, we have
\begin{equation}\label{eq:intro-expand}
  \ip{Q_uf}{Q_vf}=\sum_{\sigma\in\Omega}\omega_{uv}\tuple{\sigma}\sum_{c_u\in A_u\tuple{\sigma}}\sum_{c_v\in A_v\tuple{\sigma}}
  \tuple{f\tuple{\sigma}-f\tuple{\sigma^{u\gets c_u}}}\tuple{f\tuple{\sigma}-f\tuple{\sigma^{v\gets c_v}}}.
\end{equation}
The summand vanishes if $c_u=\sigma_u$ or $c_v=\sigma_v$.
Among the remaining terms, if $c_u\ne c_v$, the two recolorings can be performed simultaneously,
and the result $\sigma^{u\gets c_u,\,v\gets c_v}$ is again a coloring (\pref{fig:rectangle-intro}\subref{fig:rectangle-intro-different});
if $c_u=c_v$, performing both creates a conflict on $uv$ (\pref{fig:rectangle-intro}\subref{fig:rectangle-intro-same}).
Following Haqi~\cite{Haq26Sampling}, we accordingly decompose the sum into legal rectangles and pairs with a common new color.

Let $\+R_{uv}\defeq\set{(\sigma,c_u,c_v):\sigma\in\Omega,\ c_u\in A^{-\sigma_u}_u(\sigma),\ c_v\in A^{-\sigma_v}_v(\sigma),\ c_u\ne c_v}$.
Each $(\sigma,c_u,c_v)\in\+R_{uv}$ indexes a \emph{legal rectangle} with corners $\sigma$, $\sigma^{u\gets c_u}$, $\sigma^{v\gets c_v}$,
and $\sigma^{u\gets c_u,\,v\gets c_v}$, and with \emph{starting corner} $\sigma$.
The contribution of the legal rectangles to \pref{eq:intro-expand} is
\begin{equation*}
  X_{uv}\tuple{f}\defeq\sum_{\tuple{\sigma,c_u,c_v}\in\+R_{uv}}\omega_{uv}\tuple{\sigma}
  \tuple{f\tuple{\sigma}-f\tuple{\sigma^{u\gets c_u}}}\tuple{f\tuple{\sigma}-f\tuple{\sigma^{v\gets c_v}}}.
\end{equation*}
The remaining nonzero terms of \pref{eq:intro-expand} have a common new color $c_u=c_v=c\in A_u(\sigma)\cap A_v(\sigma)$;
since $u\sim v$, such a color differs from both $\sigma_u$ and $\sigma_v$.
By \pref{eq:star-def}, the term with $c_u=c_v=c$ therefore equals $\mu(\sigma)(Q_{u,c}f)(\sigma)(Q_{v,c}f)(\sigma)$,
and this product vanishes whenever $c\notin A_u(\sigma)\cap A_v(\sigma)$.
Hence
\begin{equation}\label{eq:edge-split}
  \ip{Q_uf}{Q_vf}=X_{uv}\tuple{f}+\sum_{c\in[q]}\ip{Q_{u,c}f}{Q_{v,c}f}.
\end{equation}

The following lower bound on $X_{uv}(f)$, proved below, readily implies \pref{lem:edge}.
\begin{lemma}\label{lem:legal}
  For every edge $uv\in E$ and every $f\in L^2(\mu)$,
  \begin{equation*}
    X_{uv}\tuple{f}\ge-\frac4{9\tuple{q-\Delta}^2}\tuple{\norm{Q_uf}^2+\norm{Q_vf}^2}.
  \end{equation*}
\end{lemma}

\begin{proof}[Proof of \pref{lem:edge}]
  For the lower bound, recall that $(Q_{u,c}f)(\sigma)(Q_{v,c}f)(\sigma)$ vanishes unless $c\in A_u(\sigma)\cap A_v(\sigma)$,
  in which case $c\ne\sigma_u$ and $c\ne\sigma_v$.
  Bounding each such product from below by $-\frac12((Q_{u,c}f)(\sigma)^2+(Q_{v,c}f)(\sigma)^2)$ and extending the two sums of squares to all $c\ne\sigma_u$ and all $c\ne\sigma_v$,
  respectively, we obtain
  \begin{equation*}
    \sum_{c\in[q]}\ip{Q_{u,c}f}{Q_{v,c}f}
    \ge-\frac12\sum_{\sigma\in\Omega}\mu\tuple{\sigma}\tuple{\sum_{c\ne\sigma_u}\tuple{Q_{u,c}f}\tuple{\sigma}^2
      +\sum_{c\ne\sigma_v}\tuple{Q_{v,c}f}\tuple{\sigma}^2}.
  \end{equation*}
  Since $(Q_{u,\sigma_u}f)(\sigma)=(Q_uf)(\sigma)$, the expectation of the first inner sum over $\sigma\sim\mu$ is $\sum_c\norm{Q_{u,c}f}^2-\norm{Q_uf}^2$,
  which is at most $2\norm{Q_uf}^2/(q-\Delta)$ by \pref{lem:single-vertex}.
  The second inner sum is bounded analogously, which proves the lower bound.

  For the upper bound, \pref{eq:edge-split} and \pref{lem:legal} give
  \begingroup
    \thinmuskip=2mu
    \medmuskip=2mu
    \thickmuskip=2mu
    \begin{equation*}
      \sum_{c\in[q]}\ip{Q_{u,c}f}{Q_{v,c}f}=\ip{Q_uf}{Q_vf}-X_{uv}\tuple{f}
      \le\ip{Q_uf}{Q_vf}+\frac4{9\tuple{q-\Delta}^2}\tuple{\norm{Q_uf}^2+\norm{Q_vf}^2}.\qedhere
    \end{equation*}
  \endgroup
\end{proof}

It remains to prove \pref{lem:legal}.
\begin{proof}[Proof of \pref{lem:legal}]
  Fix $(\sigma,c_u,c_v)\in\+R_{uv}$ and denote the values of $f$ at the four corners by
  \begin{equation*}
    y_{00}=f\tuple{\sigma},\qquad y_{10}=f\tuple{\sigma^{u\gets c_u}},\qquad y_{01}=f\tuple{\sigma^{v\gets c_v}},\qquad
    y_{11}=f\tuple{\sigma^{u\gets c_u,\,v\gets c_v}}.
  \end{equation*}
  Let $d_u\defeq y_{00}-y_{10}$ and $d_v\defeq y_{00}-y_{01}$ be the two first differences at $\sigma$,
  and let $M\defeq y_{00}-y_{10}-y_{01}+y_{11}$ be the mixed second difference,
  so that the summand of $X_{uv}(f)$ indexed by $(\sigma,c_u,c_v)$ is $\omega_{uv}(\sigma)\,d_ud_v$.
  The same rectangle is indexed by exactly four elements of $\+R_{uv}$, one for each of its corners taken as the starting corner;
  for example, the index with starting corner $\sigma^{u\gets c_u}$ is $(\sigma^{u\gets c_u},\sigma_u,c_v)$.

  To compare the weights of these four indices, define
  \begin{equation*}
    1+\alpha\defeq\frac{p_u\tuple{\sigma}}{p_u\tuple{\sigma^{v\gets c_v}}},\qquad
    1+\beta\defeq\frac{p_v\tuple{\sigma}}{p_v\tuple{\sigma^{u\gets c_u}}},\qquad
    w\defeq\mu\tuple{\sigma}\,p_u\tuple{\sigma^{v\gets c_v}}\,p_v\tuple{\sigma^{u\gets c_u}}.
  \end{equation*}
  Recoloring $u$ does not change the colors available at $u$, and similarly for $v$.
  Since $\mu$ is uniform, the weights and products at the four starting corners are as follows.

  \vspace{1em}
  \begin{equation*}
    \begin{array}{c@{\quad}l|c|c}
      \multicolumn{2}{c|}{\text{starting corner}} & \text{weight} & \text{product} \\[1ex]
      \rectcorner{0}{0} & \sigma
      & \omega_{uv}\tuple{\sigma}=w\tuple{1+\alpha}\tuple{1+\beta}
      & \tuple{y_{00}-y_{10}}\tuple{y_{00}-y_{01}} \\[2ex]
      \rectcorner{1}{0} & \sigma^{u\gets c_u}
      & \omega_{uv}\tuple{\sigma^{u\gets c_u}}=w\tuple{1+\alpha}
      & \tuple{y_{10}-y_{00}}\tuple{y_{10}-y_{11}} \\[2ex]
      \rectcorner{0}{1} & \sigma^{v\gets c_v}
      & \omega_{uv}\tuple{\sigma^{v\gets c_v}}=w\tuple{1+\beta}
      & \tuple{y_{01}-y_{11}}\tuple{y_{01}-y_{00}} \\[2ex]
      \rectcorner{1}{1} & \sigma^{u\gets c_u,\,v\gets c_v}
      & \omega_{uv}\tuple{\sigma^{u\gets c_u,\,v\gets c_v}}=w
      & \tuple{y_{11}-y_{01}}\tuple{y_{11}-y_{10}}
    \end{array}
  \end{equation*}
  \vspace{1em}

  Expressing the four products in terms of $d_u$, $d_v$, and $M$ (note that $y_{10}-y_{11}=d_v-M$ and $y_{01}-y_{11}=d_u-M$) and multiplying out,
  we find that the sum over the four rows of the weight times the product equals the left-hand side of the \emph{four-corner identity}
  \begin{equation}\label{eq:four-corner}
    w\tuple{M+\alpha d_u}\tuple{M+\beta d_v}
    =\frac w4\tuple{2M+\alpha d_u+\beta d_v}^2-\frac w4\tuple{\alpha d_u-\beta d_v}^2,
  \end{equation}
  whose right-hand side writes the product of the two factors as the square of half their sum minus the square of half their difference.
  With equal weights, i.e., $\alpha=\beta=0$, as for the equal-rate recolorings in~\cite{Haq26Sampling}, the four products simply add up to $M^2$.
  Since the four rows are the summands of $X_{uv}(f)$ indexed by the four indices of the same rectangle,
  summing the left-hand side of \pref{eq:four-corner} over $\+R_{uv}$ counts every summand of $X_{uv}(f)$ four times.
  Hence
  \begin{equation}\label{eq:X-rect}
    X_{uv}\tuple{f}=\frac14\sum_{\+R_{uv}}w\tuple{M+\alpha d_u}\tuple{M+\beta d_v},
  \end{equation}
  where $\sum_{\+R_{uv}}$ denotes the sum over $(\sigma,c_u,c_v)\in\+R_{uv}$,
  and $d_u$, $d_v$, $M$, $\alpha$, $\beta$, and $w$ are the quantities associated with the index.

  It remains to bound the second square in \pref{eq:four-corner}.
  Recoloring one neighbor changes the number of available colors by at most $1$, while this number is at least $q-\Delta\ge3$.
  Hence $|\alpha|,|\beta|\le1/(q-\Delta)$ and $(1+\alpha)^{-1},(1+\beta)^{-1}\le1+1/(q-\Delta)\le4/3$,
  so that $w=\omega_{uv}(\sigma)/((1+\alpha)(1+\beta))\le\frac{16}9\,\omega_{uv}(\sigma)$.
  Moreover, $d_u=f(\sigma)-f(\sigma^{u\gets c_u})$ does not depend on $c_v$, and $p_v(\sigma)|A_v(\sigma)|=1$.
  Summing out $c_v$ and using \pref{eq:first-diff}, we obtain
  \begin{equation*}
    \begin{aligned}
      \sum_{\+R_{uv}}\omega_{uv}\tuple{\sigma}\,d_u^2
      &=\sum_{\sigma\in\Omega}\mu\tuple{\sigma}\,p_u\tuple{\sigma}
      \sum_{c_u\in A_u^{-\sigma_u}\tuple{\sigma}}
      \tuple{\tuple{f\tuple{\sigma}-f\tuple{\sigma^{u\gets c_u}}}^2\,p_v\tuple{\sigma}
        \sum_{c_v\in A_v^{-\sigma_v}\tuple{\sigma},\,c_v\ne c_u}1}
      \\
      &\le\sum_{\sigma\in\Omega}\mu\tuple{\sigma}\,p_u\tuple{\sigma}
      \sum_{c_u\in A_u\tuple{\sigma}}\tuple{f\tuple{\sigma}-f\tuple{\sigma^{u\gets c_u}}}^2
      =2\norm{Q_uf}^2,
    \end{aligned}
  \end{equation*}
  and similarly $\sum_{\+R_{uv}}\omega_{uv}(\sigma)\,d_v^2\le2\norm{Q_vf}^2$.
  Combining these bounds gives
  \begin{equation}\label{eq:corner-error}
    \sum_{\+R_{uv}}w\tuple{\alpha^2d_u^2+\beta^2d_v^2}
    \le\frac{16}{9\tuple{q-\Delta}^2}\sum_{\+R_{uv}}\omega_{uv}\tuple{\sigma}\tuple{d_u^2+d_v^2}\le\frac{32}{9\tuple{q-\Delta}^2}\tuple{\norm{Q_uf}^2+\norm{Q_vf}^2}.
  \end{equation}

  Finally, we substitute \pref{eq:four-corner} into \pref{eq:X-rect}, discard the first square,
  and use $(\alpha d_u-\beta d_v)^2\le2(\alpha^2d_u^2+\beta^2d_v^2)$ together with \pref{eq:corner-error} to obtain
  \begin{align*}
    X_{uv}\tuple{f}
    &\ge-\frac1{16}\sum_{\+R_{uv}}w\tuple{\alpha d_u-\beta d_v}^2
    \ge-\frac18\sum_{\+R_{uv}}w\tuple{\alpha^2d_u^2+\beta^2d_v^2}\\
    &\ge-\frac4{9\tuple{q-\Delta}^2}\tuple{\norm{Q_uf}^2+\norm{Q_vf}^2}.\qedhere
  \end{align*}
\end{proof}

\subsection{Mixed second differences}\label{sec:mixed-squares}

We conclude this section with a bound on mixed second differences,
which reuses the analysis of the cross terms above and is needed for \pref{lem:mixed} in \pref{sec:resolvent}.
For distinct vertices $u,v\in V$, with $\omega_{uv}(\sigma)=\mu(\sigma)\,p_u(\sigma)\,p_v(\sigma)$ as in \pref{sec:adjacent}, define
\begin{equation*}
  \Theta_{uv}\tuple{f}\defeq\sum_{\sigma\in\Omega}\omega_{uv}\tuple{\sigma}
  \sum_{\substack{c_u\in A_u\tuple{\sigma},\ c_v\in A_v\tuple{\sigma}\\\sigma^{u\gets c_u,\,v\gets c_v}\in\Omega}}
  \tuple{f\tuple{\sigma}-f\tuple{\sigma^{u\gets c_u}}-f\tuple{\sigma^{v\gets c_v}}
    +f\tuple{\sigma^{u\gets c_u,\,v\gets c_v}}}^2,
\end{equation*}
and set $\Theta(f)\defeq\sum_{u\ne v}\Theta_{uv}(f)$, where the sum runs over ordered pairs of distinct vertices.
Thus $\Theta(f)$ is the weighted sum of the squared mixed second differences of $f$ over all pairs of compatible single-vertex recolorings.

\begin{lemma}\label{lem:rect}
  For every $f\in L^2(\mu)$,
  \begin{equation*}
    \Theta\tuple{f}\le16\norm{Lf}^2+\frac{38\Delta}{q-\Delta}\ip f{Lf}.
  \end{equation*}
\end{lemma}

\begin{proof}
  Fix distinct vertices $u$ and $v$, and suppose first that $u\not\sim v$.
  Then $A_v(\sigma)$ and $p_v(\sigma)$ do not depend on $\sigma_u$, and every pair of recolorings at $u$ and $v$ yields a coloring in $\Omega$.
  For fixed $\sigma_{V\setminus\{u\}}$ and $c_v\in A_v(\sigma)$,
  set $\psi(c)\defeq f(\sigma^{u\gets c})-f(\sigma^{u\gets c,\,v\gets c_v})$ for $c\in A_u(\sigma)$,
  so that the mixed second difference in $\Theta_{uv}(f)$ is $\psi(\sigma_u)-\psi(c_u)$.
  Since $A_u(\sigma^{v\gets c_v})=A_u(\sigma)$, subtracting the expectation of $\psi$ over $A_u(\sigma)$ gives $\psi(\sigma_u)-p_u(\sigma)\sum_{c\in A_u(\sigma)}\psi(c)=(Q_uf)(\sigma)-(Q_uf)(\sigma^{v\gets c_v})$.
  Applying \pref{eq:first-diff} first at $u$ (conditionally on $\sigma_{V\setminus\{u\}}$ for each fixed $c_v$) and then at $v$ (to the function $Q_uf$),
  together with \pref{eq:nonedge-positive}, gives
  \begin{align*}
    \Theta_{uv}\tuple{f}
    &=\sum_{\sigma\in\Omega}\mu\tuple{\sigma}\,p_v\tuple{\sigma}
    \sum_{c_v\in A_v\tuple{\sigma}}
    p_u\tuple{\sigma}\sum_{c_u\in A_u\tuple{\sigma}}
    \tuple{\psi\tuple{\sigma_u}-\psi\tuple{c_u}}^2\\
    &\overset{\pref{eq:first-diff}}{=}
    2\sum_{\sigma\in\Omega}\mu\tuple{\sigma}\,p_v\tuple{\sigma}
    \sum_{c_v\in A_v\tuple{\sigma}}
    \tuple{\tuple{Q_uf}\tuple{\sigma}
      -\tuple{Q_uf}\tuple{\sigma^{v\gets c_v}}}^2\\
    &\overset{\pref{eq:first-diff}}{=}
    4\norm{Q_vQ_uf}^2
    \overset{\pref{eq:nonedge-positive}}{=}4\ip{Q_uf}{Q_vf}.
  \end{align*}

  Suppose now that $uv\in E$.
  The summands of $\Theta_{uv}(f)$ with $c_u=\sigma_u$ or $c_v=\sigma_v$ vanish,
  and the remaining ones yield a proper joint recoloring if and only if $c_u\ne c_v$, as observed in \pref{sec:adjacent}.
  In the notation of the proof of \pref{lem:legal}, we therefore have $ \Theta_{uv}(f)=\sum_{(\sigma,c_u,c_v)\in\+R_{uv}}\omega_{uv}(\sigma)M^2$.
  Expanding the squares gives the algebraic identity
  \begin{equation*}
    M^2=2\tuple{M+\alpha d_u}\tuple{M+\beta d_v}+\alpha^2d_u^2+\beta^2d_v^2-\tuple{M+\alpha d_u+\beta d_v}^2.
  \end{equation*}
  Since $|\alpha|,|\beta|\le1/(q-\Delta)\le1/3$, we have $\omega_{uv}(\sigma)=w(1+\alpha)(1+\beta)\le2w$.
  Dropping the last square, multiplying by $2w$, and summing over $\+R_{uv}$, we obtain
  \begin{equation*}
    \Theta_{uv}\tuple{f}\le2\sum_{\+R_{uv}}wM^2
    \overset{\pref{eq:X-rect}, \pref{eq:corner-error}}{\le}16X_{uv}\tuple{f}+\frac{64}{9\tuple{q-\Delta}^2}\tuple{\norm{Q_uf}^2+\norm{Q_vf}^2}.
  \end{equation*}
  By \pref{eq:edge-split} and the lower bound in \pref{lem:edge}, we have $X_{uv}(f)\le\ip{Q_uf}{Q_vf}+(\norm{Q_uf}^2+\norm{Q_vf}^2)/(q-\Delta)$.
  Using $q-\Delta\ge3$ to bound $64/(9(q-\Delta)^2)\le3/(q-\Delta)$, we obtain
  \begin{equation*}
    \Theta_{uv}\tuple{f}\le16\ip{Q_uf}{Q_vf}+\frac{19}{q-\Delta}\tuple{\norm{Q_uf}^2+\norm{Q_vf}^2}.
  \end{equation*}

  Finally, we sum the two bounds over all ordered pairs $u\ne v$.
  Since $\ip{Q_uf}{Q_vf}=\norm{Q_uQ_vf}^2\ge0$ for $u\not\sim v$, the non-adjacent bound implies $\Theta_{uv}(f)\le16\ip{Q_uf}{Q_vf}$ as well.
  Combining both cases with \pref{eq:expand-L} and \pref{eq:edge-count} gives
  \begin{align*}
    \Theta\tuple{f}
    &\le16\sum_{u\ne v}\ip{Q_uf}{Q_vf}+\frac{38\Delta}{q-\Delta}\ip f{Lf}\\
    &=16\tuple{\norm{Lf}^2-\ip f{Lf}}+\frac{38\Delta}{q-\Delta}\ip f{Lf}
    \le16\norm{Lf}^2+\frac{38\Delta}{q-\Delta}\ip f{Lf}.\qedhere
  \end{align*}
\end{proof}

\section{The hidden hard-core model}\label{sec:hidden}

In this section, we prove \pref{lem:hardcore-comparison}.
We split $\ip f{L_cf}$ into a mean part and a variance part in \pref{sec:split},
construct the hard-core model hidden in the color class of $c$ in \pref{sec:ghost},
and relate its dynamics to $L_c$ by averaging over the ghost colors in \pref{sec:surviving}.
In \pref{sec:two-bounds}, we smooth $f$ by the resolvents of the hidden dynamics and deduce \pref{lem:hardcore-comparison} from two estimates proved in \pref{sec:resolvent}.
Throughout this section, unless stated otherwise, we fix a color $c\in[q]$.

\subsection{Separating the mean and variance parts}\label{sec:split}

The single-color operators need not individually satisfy an integrated Bochner inequality with a constant independent of $q$.
To see this, let $v$ be an isolated vertex, so that $A_v(\sigma)=[q]$ for every coloring $\sigma$, and consider two test functions $f_1$ and $f_2$.
For $f_1(\sigma)\defeq\phi(\sigma_v)$, where $\phi\colon[q]\to\^R$ is nonzero and satisfies $\phi(c)=0$ and $\sum_{c'\ne c}\phi(c')=0$,
\pref{eq:star-def} gives $L_cf_1=f_1/q$, and hence $\norm{L_cf_1}^2=\ip{f_1}{L_cf_1}/q$.
For $f_2(\sigma)\defeq\one\{\sigma_v=c\}-1/q$, the same identity gives $L_cf_2=f_2$.
Hence $\norm{L_cf}^2\ge\kappa\ip f{L_cf}$ can hold for all $f$ only if $\kappa\le1/q$.
More generally, fix a vertex $v$ and the coloring $\sigma_{V\setminus\{v\}}$ with $c\in A_v(\sigma)$.
As $\sigma_v$ changes, we distinguish two kinds of variation in $f$.
\begin{enumerate}[label=\textup{(\roman*)},nosep]
  \item the difference between the value of $f$ at color $c$ and its expectation over the other available colors;
  \item the variation in the values of $f$ among those other colors.
\end{enumerate}
In the example above, $f_2$ has only the first kind of variation and $f_1$ only the second.
A general function may have both, so we separate their contributions to $\ip f{L_cf}$.

For $f\colon\Omega\to\^R$ and $v\in V$, define the functions $\mathbb{E}_v^{-c}f$ and $\operatorname{Var}_v^{-c}f$ on $\Omega$ by
\begin{align*}
  \mathbb{E}_v^{-c}f\tuple{\sigma}\defeq\frac1{|A^{-c}_v\tuple{\sigma}|}\sum_{c'\in A^{-c}_v\tuple{\sigma}}f\tuple{\sigma^{v\gets c'}},\qquad
  \operatorname{Var}_v^{-c}f\defeq\mathbb{E}_v^{-c}\tuple{f^2}-\tuple{\mathbb{E}_v^{-c}f}^2.
\end{align*}
For $f\in L^2(\mu)$, the above two kinds of variation are captured by
\begin{align*}
  \+M_{v,c}\tuple{f}&\defeq\E[\sigma\sim\mu]{\one\{\sigma_v=c\}\tuple{1-p_v}\tuple{f-\mathbb{E}_v^{-c}f}^2},\\
  \+V_{v,c}\tuple{f}&\defeq\E[\sigma\sim\mu]{\one\{\sigma_v=c\}\tuple{1-p_v}\operatorname{Var}_v^{-c}f}.
\end{align*}
Equivalently, $\+M_{v,c}(f)$ is the expectation over $\sigma_{V\setminus\{v\}}$ of the squared norm of the orthogonal projection of $f$ onto the span of the centered indicator of color $c$ at $v$ in the conditional $L^2$ space.
Define the \emph{mean part} $\+M_c\defeq\sum_v\+M_{v,c}$ and the \emph{variance part} $\+V_c\defeq\sum_v\+V_{v,c}$.

\begin{lemma}\label{lem:MV-split}
  For every $f\in L^2(\mu)$, $\ip f{L_cf}=\+M_c(f)+\+V_c(f)$.
\end{lemma}

\begin{proof}
  By reversibility,
  \begin{align}
    \ip f{Q_{v,c}f}
    &=\E[\sigma\sim\mu]{\one\{\sigma_v=c\}\,p_v\tuple{\sigma}
      \sum_{c'\in A^{-c}_v\tuple{\sigma}}\tuple{f\tuple{\sigma}-f\tuple{\sigma^{v\gets c'}}}^2}\notag\\
    &=\E[\sigma\sim\mu]{\one\{\sigma_v=c\}\tuple{1-p_v\tuple{\sigma}}
      \tuple{f^2\tuple{\sigma}-2f\tuple{\sigma}\mathbb{E}_v^{-c}f\tuple{\sigma}
        +\mathbb{E}_v^{-c}\tuple{f^2}\tuple{\sigma}}}\notag\\
    &=\+M_{v,c}\tuple{f}+\+V_{v,c}\tuple{f},\notag
  \end{align}
  where the second equality uses $p_v(\sigma)|A^{-c}_v(\sigma)|=1-p_v(\sigma)$ when $\sigma_v=c$.
  Summing over $v$ and using $L_c=\sum_vQ_{v,c}$ proves the claim.
\end{proof}

For a fixed color $c$, the mean part $\+M_c(f)$ need not control the single-color Dirichlet form $\ip f{L_cf}$.
The following lemma shows that, after summing over all colors, the mean parts nevertheless control the full Dirichlet form $\ip f{Lf}$.

\begin{lemma}\label{lem:mean-sum}
  For every $f\in L^2(\mu)$, we have $\sum_{c\in[q]}\+M_c(f)\ge\ip f{Lf}$ and $\sum_{c\in[q]}\+V_c(f)\le\ip f{Lf}$.
\end{lemma}

\begin{proof}
  If $\sigma_v=c$, then $p_v(\sigma)|A^{-c}_v(\sigma)|=1-p_v(\sigma)$, so the first case of \pref{eq:star-def} gives
  \begin{equation}\label{eq:Qv-mean-diff}
    \tuple{Q_{v,\sigma_v}f}\tuple{\sigma}=\tuple{Q_vf}\tuple{\sigma}
    =\tuple{1-p_v\tuple{\sigma}}\tuple{f\tuple{\sigma}-\mathbb{E}_v^{-\sigma_v}f\tuple{\sigma}}.
  \end{equation}
  Summing over $c\in[q]$ and $v\in V$ and using $1-p_v(\sigma)\le1$, we obtain
  \begin{align*}
    \sum_{c\in[q]}\+M_c\tuple{f}
    =\sum_{v\in V}\E[\sigma\sim\mu]{\frac{\tuple{\tuple{Q_vf}\tuple{\sigma}}^2}{1-p_v\tuple{\sigma}}}
    \ge\sum_{v\in V}\norm{Q_vf}^2=\ip f{Lf}.
  \end{align*}
  Consequently, by \pref{lem:star} and \pref{lem:MV-split}, $\sum_c\+V_c(f)=2\ip f{Lf}-\sum_c\+M_c(f)\le\ip f{Lf}$.
\end{proof}

We will also use the bilinear forms associated with $\+M_c$ and $\+V_c$.
Since $\+M_c$ and $\+V_c$ are nonnegative quadratic forms on $L^2(\mu)$, their polarizations
\begin{equation*}
  \+M_c\tuple{f,g}\defeq\frac14\tuple{\+M_c\tuple{f+g}-\+M_c\tuple{f-g}},\qquad
  \+V_c\tuple{f,g}\defeq\frac14\tuple{\+V_c\tuple{f+g}-\+V_c\tuple{f-g}}
\end{equation*}
are symmetric positive semidefinite bilinear forms with $\+M_c(f,f)=\+M_c(f)$ and $\+V_c(f,f)=\+V_c(f)$.
They retain the weights in $\+M_c$ and $\+V_c$, with $(f-\mathbb{E}_v^{-c}f)^2$ and $\operatorname{Var}_v^{-c}f$ replaced, respectively, by
\begin{equation*}
  \tuple{f-\mathbb{E}_v^{-c}f}\tuple{g-\mathbb{E}_v^{-c}g},
  \qquad
  \mathbb{E}_v^{-c}\tuple{fg}-\tuple{\mathbb{E}_v^{-c}f}\tuple{\mathbb{E}_v^{-c}g}.
\end{equation*}
Since $L_c$ is self-adjoint, polarizing \pref{lem:MV-split} gives $\ip f{L_cg}=\+M_c(f,g)+\+V_c(f,g)$.

\subsection{Hidden hard-core model}\label{sec:ghost}

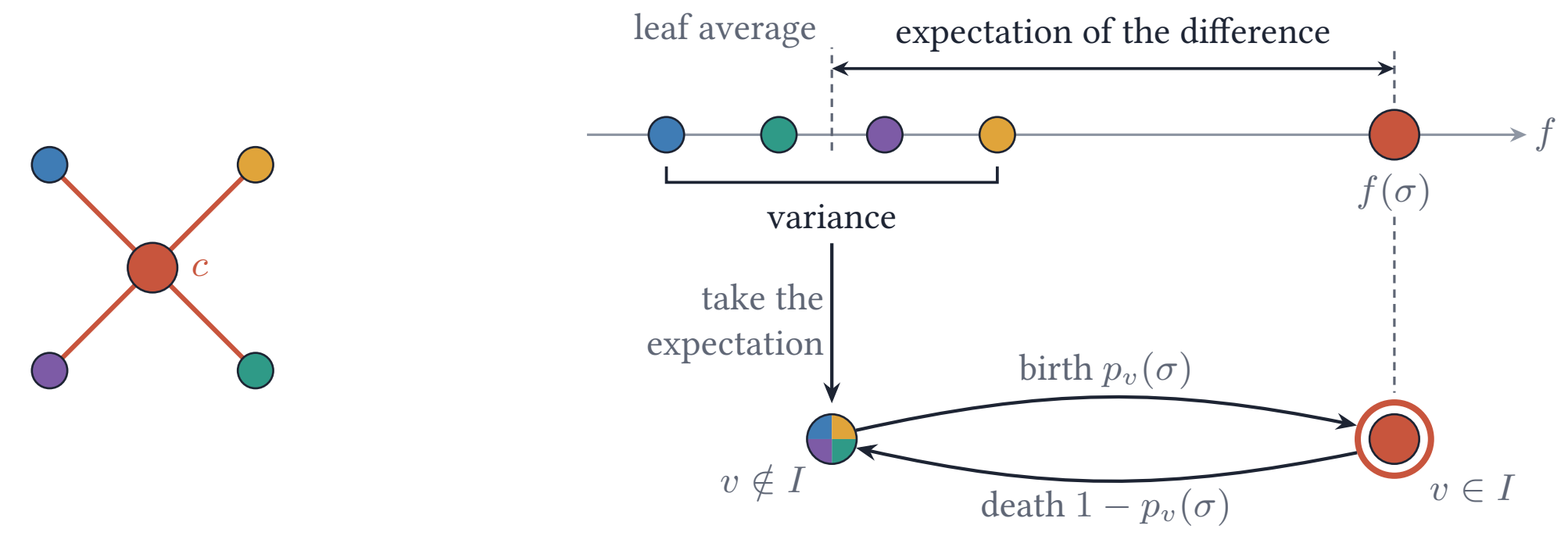
\begin{figure}[H]
  \centering
  \begin{subfigure}[t]{0.28\linewidth}
    \centering
    \begin{tikzpicture}[x=1.12cm,y=1.12cm]
      \path[use as bounding box] (-1.25,-3.1) rectangle (1.25,0.9);
      \begin{scope}[shift={(0,-1.1)}]
        \foreach \a/\col in {135/gblue,45/ggold,315/gteal,225/gpurple}{\draw[gx,line width=1.15pt] (0,0) -- (\a:1.1);
          \node[vert,fill=\col] at (\a:1.1) {};
        }
        \node[vert,fill=gx,minimum size=0.42cm] at (0,0) {};
        \node[flab,text=gx,anchor=west] at (0.24,0) {$c$};
      \end{scope}
    \end{tikzpicture}
    \caption{Recolorings involving $c$.}
    \label{fig:mean-variance-star}
  \end{subfigure}\hspace{0.03\linewidth}\begin{subfigure}[t]{0.64\linewidth}
    \centering
    \begin{tikzpicture}[x=1.12cm,y=1.12cm]
      \draw[cedge,line width=0.6pt,-{Stealth[length=4pt,width=3.5pt]}] (2.3,0) -- (9.4,0)
        node[right,flab] {$f$};
      \draw[cmuted,line width=0.6pt,dash pattern=on 2pt off 1.5pt] (4.15,-0.12) -- (4.15,0.66);
      \draw[cmuted,line width=0.6pt,dash pattern=on 2pt off 1.5pt] (8.4,0.24) -- (8.4,0.66);
      \draw[{Stealth[length=4pt,width=3.5pt]}-{Stealth[length=4pt,width=3.5pt]},cink,line width=0.7pt]
        (4.15,0.5) -- node[above,fcap] {expectation of the difference} (8.4,0.5);
      \node[flab,anchor=south east] at (4.09,0.6) {leaf average};
      \draw[cink,line width=0.7pt] (2.9,-0.24) -- (2.9,-0.36) -- (5.4,-0.36) -- (5.4,-0.24);
      \node[fcap,anchor=north] at (4.15,-0.4) {variance};
      \foreach \x/\col in {2.9/gblue,3.75/gteal,4.55/gpurple,5.4/ggold}{\node[vert,fill=\col] at (\x,0) {};
      }
      \node[vert,fill=gx,minimum size=0.42cm] at (8.4,0) {};
      \node[flab,anchor=north] at (8.4,-0.24) {$f(\sigma)$};
      \draw[move] (4.15,-0.82) -- node[left,flab,align=right] {take the\\expectation} (4.15,-2.03);
      \draw[cmuted,line width=0.6pt,dash pattern=on 2pt off 1.5pt] (8.4,-0.62) -- (8.4,-1.95);
      \foreach \a/\col in {90/gblue,0/ggold,270/gteal,180/gpurple}{\fill[\col] (4.15,-2.3) -- ++(\a:0.21cm) arc[start angle=\a,delta angle=90,radius=0.21cm] -- cycle;
      }
      \draw[cink,line width=0.5pt] (4.15,-2.3) circle[radius=0.21cm];
      \node[circle,minimum size=0.42cm,inner sep=0pt] (vout) at (4.15,-2.3) {};
      \node[ring,minimum size=0.62cm] (vin) at (8.4,-2.3) {};
      \node[vert,fill=gx,minimum size=0.42cm] at (8.4,-2.3) {};
      \draw[move] (vout.20) to[bend left=12] node[above,flab] {birth $p_v(\sigma)$} (vin.160);
      \draw[move] (vin.200) to[bend left=12] node[below,flab] {death $1-p_v(\sigma)$} (vout.340);
      \node[flab,anchor=north east] at (vout.south west) {$v\notin I$};
      \node[flab,anchor=north west] at (vin.south east) {$v\in I$};
    \end{tikzpicture}
    \caption{Values of $f$ and the hard-core update.}
    \label{fig:mean-variance-values}
  \end{subfigure}
  \caption[The mean and variance parts]{The mean and variance parts at a vertex $v$ with $\sigma_v=c$.
    (a) The moves of $Q_{v,c}$ form the star centered at $c$, as in \pref{fig:stars}.
    (b) The top shows the values of $f$ at the four leaves and at the center.
    The mean part $\+M_{v,c}(f)$ measures the difference between $f(\sigma)$ and the average leaf value,
    and the variance part $\+V_{v,c}(f)$ measures the spread of the leaf values.
    The bottom shows the two-state hard-core update at $v$, with birth probability $p_v(\sigma)$ and death probability $1-p_v(\sigma)$.}
  \label{fig:mean-variance}
\end{figure}

Each summand $Q_{v,c}$ of $L_c$ acts along the star centered at $c$ (\pref{fig:mean-variance}\subref{fig:mean-variance-star}).
The corresponding update $P_{v,c}$ changes the color class $I=\{v\in V:\sigma_v=c\}$ through births and deaths (\pref{fig:mean-variance}\subref{fig:mean-variance-values}).
An unblocked vertex $v\notin I$ joins $I$ with probability $p_v(\sigma)$,
whereas a vertex $v\in I$ leaves $I$ with probability $1-p_v(\sigma)$ by taking a uniformly random color from $A^{-c}_v(\sigma)$.
Because $p_v(\sigma)$ can change when a neighbor of $v$ leaves $I$ and acquires a new color,
we sample in advance the color that each vertex of $I$ takes upon leaving.

Specifically, let $\Omega^{-c}\defeq\{\eta\in\Omega:\eta_v\ne c\text{ for all }v\in V\}$, and call an element of $\Omega^{-c}\times\+I$ a \emph{pair}.
To represent a coloring $\sigma\in\Omega$ by a pair $(\eta,I)$, set $I=\{v\in V:\sigma_v=c\}$,
choose $\eta_v\in A^{-c}_v(\sigma)$ independently and uniformly for each $v\in I$, and set $\eta_v=\sigma_v$ for $v\notin I$.
Since $I$ is an independent set, no two recolored vertices are adjacent, so $\eta\in\Omega^{-c}$.
We call the resulting random pair $(\eta,I)$ a \emph{lift} of $\sigma$,
call $(\eta_v)_{v\in I}$ the \emph{ghost colors} and $\eta$ the \emph{background coloring},
and write $\muh_c$ for the distribution of a lift of $\sigma\sim\mu$.
In the hidden dynamics defined below, $\eta$ stays fixed while $I$ evolves.

Conversely, for any pair $(\eta,I)\in\Omega^{-c}\times\+I$, let $\pi_c(\eta,I)\in\Omega$ be the coloring that equals $c$ on $I$ and agrees with $\eta$ outside $I$.
Since $\pi_c(\eta,I)=\sigma$ for every lift $(\eta,I)$ of $\sigma$, if $(\eta,I)\sim\muh_c$, then $\pi_c(\eta,I)\sim\mu$,
and conditionally on $\pi_c(\eta,I)=\sigma$, the pair $(\eta,I)$ is a lift of $\sigma$.
Moreover, for every $v\in I$, the neighbors of $v$ lie outside $I$, so $A_v(\pi_c(\eta,I))=A_v(\eta)$.
For each $\eta\in\Omega^{-c}$ and $v\in V$, define the vertex activity
\begin{equation*}
  \lambda_v\tuple{\eta}\defeq\frac1{|A_v\tuple{\eta}|-1}\le\frac1{q-\Delta-1}.
\end{equation*}
For each $\eta\in\Omega^{-c}$, let $\nu_\eta$ denote the hard-core model on $G$ with activities $\lambda(\eta)\defeq(\lambda_v(\eta))_{v\in V}$, i.e.,
the distribution $\nu$ from \pref{sec:prelim-hardcore} with $\lambda=\lambda(\eta)$;
we call $\nu_\eta$ the \emph{hidden model}.

\begin{lemma}\label{lem:lift}
  Let $(\eta,I)\sim\muh_c$.
  Conditioned on $\eta$, the independent set $I$ has distribution $\nu_\eta$.
\end{lemma}

\begin{proof}
  Fix $(\eta,I)\in\Omega^{-c}\times\+I$ and write $\sigma\defeq\pi_c(\eta,I)$.
  A lift produces $(\eta,I)$ if and only if the initial coloring is $\sigma$ and the ghost color chosen at each $v\in I$ is $\eta_v$.
  For every $v\in I$, we have $A_v(\sigma)=A_v(\eta)$ and hence $\eta_v\in A^{-c}_v(\sigma)$.
  Since the ghost colors are chosen independently and uniformly,
  \begin{equation}\label{eq:lift-distribution}
    \muh_c\tuple{\eta,I}
    =\mu\tuple{\sigma}
    \prod_{v\in I}\frac1{|A_v\tuple{\sigma}|-1}
    =\frac1{|\Omega|}\prod_{v\in I}\lambda_v\tuple{\eta}.
  \end{equation}
  For fixed $\eta\in\Omega^{-c}$, every $I\in\+I$ has positive probability under $\muh_c$,
  and normalizing \pref{eq:lift-distribution} over $I\in\+I$ yields $\nu_\eta(I)$ by \pref{eq:hardcore}.
\end{proof}

\paragraph{The hidden dynamics and its spectral gap.}
For each background coloring $\eta\in\Omega^{-c}$, we call the Glauber dynamics for $\nu_\eta$ the \emph{hidden dynamics},
and write $H_{\eta,v}$ and $H_\eta$ for the operators $H_v$ and $H$ of \pref{sec:prelim-hardcore} with activities $\lambda=\lambda(\eta)$.
Under $\nu_\eta$, an unblocked vertex $v$ is occupied with conditional probability $\lambda_v(\eta)/(1+\lambda_v(\eta))=p_v(\eta)$.
Whenever $q\ge(1+1/\ee+\gamma)\Delta$ and $\Delta$ is sufficiently large,
the upper bound $\lambda_v(\eta)\le 1/(q-\Delta-1)$ lies uniformly below the uniqueness threshold $\lambda_{\mathrm{crit}}(\Delta)\sim \ee/\Delta$,
so \pref{thm:hardcore} yields a uniform spectral gap for all hidden models simultaneously.

\begin{corollary}\label{cor:lifted-gap}
  For every $\gamma>0$, let $\theta_\gamma\defeq\theta_*(\delta_\gamma)\in(0,1]$,
  where $\theta_*$ is the function from \pref{thm:hardcore} and $\delta_\gamma\defeq(\ee\gamma/2)/(1+\ee\gamma/2)$.
  If $q\ge(1+1/\ee+\gamma)\Delta$ and $\Delta\ge\max\{3,2/\gamma\}$, then $\gap(H_\eta)\ge\theta_\gamma$ for every $c\in[q]$ and every $\eta\in\Omega^{-c}$.
\end{corollary}

\begin{proof}
  Since $q\ge(1+1/\ee+\gamma)\Delta$ and $\gamma\Delta\ge2$, we have $q-\Delta-1\ge(1/\ee+\gamma)\Delta-1\ge(1/\ee+\gamma/2)\Delta$.
  Moreover, $(1+1/m)^{m+1}>\ee$ for every $m\ge1$, so
  \begin{equation*}
    \lambda_{\mathrm{crit}}\tuple{\Delta}=\frac1{\Delta-2}\tuple{1+\frac1{\Delta-2}}^{\Delta-1}>\frac \ee{\Delta-2}>\frac \ee\Delta.
  \end{equation*}
  Consequently, for every $v\in V$,
  \begin{equation*}
    \lambda_v\tuple{\eta}\le\frac1{q-\Delta-1}\le\frac1{\tuple{\frac{1}{\ee}+\frac{\gamma}{2}}\Delta}<\frac{\lambda_{\mathrm{crit}}\tuple{\Delta}}{1+\frac{\ee\gamma}{2}}=\tuple{1-\delta_\gamma}\lambda_{\mathrm{crit}}\tuple{\Delta},
  \end{equation*}
  and the claim follows from \pref{thm:hardcore} with $\delta=\delta_\gamma$.
\end{proof}

\subsection{Lifting and averaging over ghost colors}\label{sec:surviving}

To relate the hidden dynamics to the operator $L_c$ on $\Omega$, we use lifting and averaging.
For $f\colon\Omega\to\^R$ and $F\colon\Omega^{-c}\times\+I\to\^R$,
define the \emph{lift} $\lift f$ of $f$ and the \emph{average} $\down F$ of $F$ over the ghost colors by
\begin{equation*}
  \lift f\tuple{\eta,I}\defeq f\tuple{\pi_c\tuple{\eta,I}},\qquad
  \down F\tuple{\sigma}\defeq\E[\tuple{\eta,I}\sim\muh_c]{F\tuple{\eta,I}\mid\pi_c\tuple{\eta,I}=\sigma}.
\end{equation*}
Thus $\lift f$ takes the value $f(\sigma)$ at every lift of $\sigma$, and $\down F(\sigma)$ is the expectation of $F$ at a random lift of $\sigma$.
We suppress the dependence of $\lift f$ and $\down F$ on $c$.
For $F\colon\Omega^{-c}\times\+I\to\^R$ and $\eta\in\Omega^{-c}$, we write $F_\eta\defeq F(\eta,\cdot)$, a function on $\+I$.
In particular, $\lift f_\eta(I)=f(\pi_c(\eta,I))$, and $\down{(\lift f)}=f$.
Write $\rho_c$ for the marginal distribution of the background coloring, so that $\muh_c(\eta,I)=\rho_c(\eta)\nu_\eta(I)$ by \pref{lem:lift}.
Since $\pi_c(\eta,I)$ has distribution $\mu$ when $(\eta,I)\sim\muh_c$,
conditioning on $\pi_c(\eta,I)$ shows that $\ip[\muh_c]{\lift f}F=\ip f{\down F}$ for all $f\in L^2(\mu)$ and $F\in L^2(\muh_c)$,
and Jensen's inequality for conditional expectations gives the contraction bound
\begin{equation}\label{eq:averaging-contraction}
  \norm{\down F}\le\norm[\muh_c]F.
\end{equation}

\begin{lemma}\phantomsection\label{lem:compression}
  \leavevmode
  \begin{enumerate}[label=\textup{(\alph*)}]
    \item\label{lem:compression-a} For every $f\in L^2(\mu)$,
      \begin{equation}\label{eq:lift-expectation}
        \E[\eta\sim\rho_c]{\ip[\nu_\eta]{\lift f_\eta}{H_\eta\lift f_\eta}}=\ip f{L_cf}.
      \end{equation}
    \item\label{lem:compression-b} For every function $F\in L^2(\muh_c)$,
      \begin{equation}\label{eq:compression-b}
        \E[\eta\sim\rho_c]{\ip[\nu_\eta]{F_\eta}{H_\eta F_\eta}}\ge\+M_c\tuple{\down F}.
      \end{equation}
  \end{enumerate}
\end{lemma}

\begin{proof}
  Let $(\eta,I)\sim\muh_c$ and write $\sigma\defeq\pi_c(\eta,I)\sim\mu$.
  For each $v\in V$, we have $v\in I$ if and only if $\sigma_v=c$, in which case $p_v(\eta)=p_v(\sigma)$.
  Since $I\sim\nu_\eta$ conditionally on $\eta$ by \pref{lem:lift}, summing \pref{eq:hc-death} over $v\in V$ and conditioning on $\sigma$ gives,
  for every $F\in L^2(\muh_c)$,
  \begin{equation}\label{eq:death-conditioned}
    \begin{aligned}
      &\E[\eta\sim\rho_c]{\ip[\nu_\eta]{F_\eta}{H_\eta F_\eta}}\\
      &\qquad=\sum_{v\in V}\E[\tuple{\eta,I}\sim\muh_c]{\one\{v\in I\}\tuple{1-p_v\tuple{\sigma}}
        \tuple{F\tuple{\eta,I}-F\tuple{\eta,I\setminus\{v\}}}^2}\\
      &\qquad=\sum_{v\in V}\E[\sigma\sim\mu]{\one\{\sigma_v=c\}\tuple{1-p_v\tuple{\sigma}}
        \E[\tuple{\eta,I}\sim\muh_c]{\tuple{F\tuple{\eta,I}-F\tuple{\eta,I\setminus\{v\}}}^2\mid\pi_c\tuple{\eta,I}=\sigma}}.
    \end{aligned}
  \end{equation}

  Fix $v\in V$ and $\sigma\in\Omega$ with $\sigma_v=c$.
  Conditionally on $\pi_c(\eta,I)=\sigma$, the ghost color $\eta_v$ is uniform on $A^{-c}_v(\sigma)$.
  For each $c'\in A^{-c}_v(\sigma)$, recoloring $v$ with $c'$ removes $v$ from the $c$-class $I$ and leaves the sets of non-$c$ colors available at the vertices of $I\setminus\{v\}$ unchanged,
  since none of them is adjacent to $v$.
  Thus, conditioned on $\pi_c(\eta,I)=\sigma$ and $\eta_v=c'$, the pair $(\eta,I\setminus\{v\})$ is a random lift of $\sigma^{v\gets c'}$.

  (a) Take $F=\lift f$.
  Since $\pi_c(\eta,I\setminus\{v\})=\sigma^{v\gets\eta_v}$, the inner conditional expectation in \pref{eq:death-conditioned} is the expectation over $\eta_v\in A^{-c}_v(\sigma)$ of $(f(\sigma)-f(\sigma^{v\gets\eta_v}))^2$,
  which equals $f^2(\sigma)-2f(\sigma)\mathbb{E}_v^{-c}f(\sigma)+\mathbb{E}_v^{-c}(f^2)(\sigma)$.
  Substituting this identity into \pref{eq:death-conditioned} gives $\+M_c(f)+\+V_c(f)=\ip f{L_cf}$ by \pref{lem:MV-split}.

  (b) For a general $F\in L^2(\muh_c)$, taking the conditional expectation given $\pi_c(\eta,I)=\sigma$ and $\eta_v=c'$ yields $\E[(\eta,I)\sim\muh_c]{F(\eta,I\setminus\{v\})\mid\pi_c(\eta,I)=\sigma,\ \eta_v=c'}=\down F(\sigma^{v\gets c'})$.
  Taking the expectation over $\eta_v$ and using $\E[(\eta,I)\sim\muh_c]{F(\eta,I)\mid\pi_c(\eta,I)=\sigma}=\down F(\sigma)$ gives
  \begin{equation*}
    \E[\tuple{\eta,I}\sim\muh_c]{F\tuple{\eta,I}-F\tuple{\eta,I\setminus\{v\}}\mid\pi_c\tuple{\eta,I}=\sigma}
    =\down F\tuple{\sigma}-\mathbb{E}_v^{-c}\down F\tuple{\sigma}.
  \end{equation*}
  Bounding the conditional expectation of the square in \pref{eq:death-conditioned} from below by the square of this conditional expectation via Jensen's inequality yields \hyperref[lem:compression-b]{(b)}.
\end{proof}

\subsection{Smoothing by resolvents: Proof of \texorpdfstring{\pref{lem:hardcore-comparison}}{Lemma \getrefnumber{lem:hardcore-comparison}}}\label{sec:two-bounds}

Since the hidden dynamics keeps $\eta$ fixed, it leaves unchanged every function that depends only on $\eta$.
After taking the expectation over the ghost colors, such a function need not be constant on $\Omega$.
This explains why the spectral gaps for fixed backgrounds do not directly give a spectral gap for $L_c$.
Instead, we use these gaps to control the forms $\+M_c$ by smoothing $f$ through the \emph{resolvents} of the hidden dynamics.

Fix $a>0$ and $f\in L^2(\mu)$.
For each color $c\in[q]$ and background coloring $\eta\in\Omega^{-c}$, we smooth $\lift f_\eta$ by minimizing
\begin{equation*}
  \norm[\nu_\eta]{g-\lift f_\eta}^2
  +a\ip[\nu_\eta]g{H_\eta g}
  \qquad\text{over }g\in L^2(\nu_\eta).
\end{equation*}
The first term penalizes the displacement from $\lift f_\eta$, while the second penalizes variation along the hidden updates.
The parameter $a$ determines the relative weight of these two costs.
Since $H_\eta$ is self-adjoint and positive semidefinite, this objective is strictly convex, and its unique minimizer satisfies $(\Id+aH_\eta)g=\lift f_\eta$.
Thus the minimizer is $R_\eta\lift f_\eta$, where $R_\eta\defeq(\Id+aH_\eta)^{-1}$ is the resolvent of the hidden dynamics.

Let $\widehat R_c\colon L^2(\muh_c)\to L^2(\muh_c)$ be defined by $(\widehat R_cF)_\eta\defeq R_\eta F_\eta$ for each fixed background coloring $\eta$,
and define the smoothed version of $f$ on $\Omega$ by
\begin{equation*}
  \widetilde f_c\defeq\down{\tuple{\widehat R_c\lift f}}.
\end{equation*}
We suppress the dependence on $a$ in $R_\eta$, $\widehat R_c$, and $\widetilde f_c$.

\begin{remark}[Probabilistic interpretation of $\widetilde f_c$]
  In terms of the transition matrix $P_\eta\defeq P^{\nu_\eta}=\Id-H_\eta/n$ of the hidden dynamics, the resolvent expands as
  \begin{equation*}
    R_\eta=\frac1{1+an}\tuple{\Id-\frac{an}{1+an}P_\eta}^{-1}
    =\sum_{k=0}^{\infty}\frac{\tuple{an}^k}{\tuple{1+an}^{k+1}}\tuple{P_\eta}^k.
  \end{equation*}
  Thus $(R_\eta h)(I)=\E{h(I_N)\mid I_0=I}$, where $(I_k)_{k\ge0}$ is the hidden dynamics, i.e., the chain with transition matrix $P_\eta$,
  and $N$ is independent of $(I_k)_{k\ge0}$ with $\Pr{N=k}=(an)^k/(1+an)^{k+1}$ for $k\ge0$;
  in particular, $\E{N}=an$.
  Combined with lifting and averaging, this shows that $\widetilde f_c(\sigma)$ is the expected output of the following procedure:
  \begin{enumerate}[label=\textup{(\arabic*)},nosep]
    \item sample a random lift $(\eta,I)$ of $\sigma$;
    \item sample $N$ as above, and run the hidden dynamics from $I_0=I$ for $N$ steps with $\eta$ fixed;
    \item output $f(\pi_c(\eta,I_N))$.
  \end{enumerate}
\end{remark}

Recalling the bilinear forms defined in \pref{sec:split} and writing $\+M_c(f)=\+M_c(f,f)$, we obtain
\begin{equation}\label{eq:smoothing-split}
  \+M_c\tuple{f}=\+M_c\tuple{f-\widetilde f_c,f}+\+M_c\tuple{\widetilde f_c,f}
  =\ip{f-\widetilde f_c}{L_cf}-\+V_c\tuple{f-\widetilde f_c,f}+\+M_c\tuple{\widetilde f_c,f}.
\end{equation}
After summing over $c$, we bound these three terms using the Cauchy--Schwarz inequality and Young's inequality.
\pref{lem:mean}\hyperref[lem:mean-a]{(a)} controls $\sum_c\norm{f-\widetilde f_c}^2$, which appears in the inner-product term;
for the $\+V_c$ cross term, \pref{lem:variance} bounds $\sum_c\+V_c(f-\widetilde f_c)$;
and \pref{lem:mean}\hyperref[lem:mean-b]{(b)} bounds $\sum_c\+M_c(\widetilde f_c)$,
which allows the $\+M_c$ cross term to be absorbed for a suitable choice of $a$.
Both lemmas are proved in \pref{sec:resolvent}.

\begin{lemma}\label{lem:mean}
  Let $a>0$.
  For every $f\in L^2(\mu)$, the following hold.
  \begin{enumerate}[label=\textup{(\alph*)}]
    \item\label{lem:mean-a}
      \begin{equation*}
        \sum_{c\in[q]}\norm{f-\widetilde f_c}^2\le2a\ip f{Lf}.
      \end{equation*}
    \item\label{lem:mean-b} If $\theta>0$ and $\gap(H_\eta)\ge\theta$ for every $c\in[q]$ and every $\eta\in\Omega^{-c}$, then
      \begin{equation*}
        \sum_{c\in[q]}\+M_c\tuple{\widetilde f_c}\le\frac2{\tuple{1+a\theta}^2}\ip f{Lf}.
      \end{equation*}
  \end{enumerate}
\end{lemma}

\begin{lemma}\label{lem:variance}
  Suppose that $q\ge(1+1/\ee)\Delta$.
  There is an absolute constant $C_0$ such that, for every $a>0$ and every $f\in L^2(\mu)$,
  \begin{equation*}
    \sum_{c\in[q]}\+V_c\tuple{f-\widetilde f_c}\le\frac{C_0a}{q-\Delta}\tuple{\norm{Lf}^2+\ip f{Lf}}+\frac{C_0a^2}{\tuple{q-\Delta}^2}\ip f{Lf}.
  \end{equation*}
\end{lemma}

Together with \pref{lem:mean-sum} and \pref{cor:lifted-gap}, these estimates imply \pref{lem:hardcore-comparison}.

\begin{proof}[Proof of \pref{lem:hardcore-comparison}]
  Let $\theta_\gamma$ be the constant from \pref{cor:lifted-gap}, and take $a=3/\theta_\gamma$, so that $2/(1+a\theta_\gamma)^2=1/8$.
  The assumptions of the lemma imply $q\ge(1+1/\ee)\Delta$, so \pref{lem:variance} applies;
  moreover, by \pref{cor:lifted-gap}, the hidden dynamics satisfies $\gap(H_\eta)\ge\theta_\gamma$ for every $c$ and $\eta$.
  Fix $f\in L^2(\mu)$.

  We sum \pref{eq:smoothing-split} over $c$ and bound the three resulting sums by the Cauchy--Schwarz inequality,
  applied to the inner product on $L^2(\mu)$ and to the positive semidefinite forms $\+V_c$ and $\+M_c$, respectively.
  This gives
  \begin{align*}
    \sum_{c\in[q]}\ip{f-\widetilde f_c}{L_cf}
    &\le \tuple{\sum_{c\in[q]} \norm{f-\widetilde f_c}^2}^{\frac12}\tuple{\sum_{c\in[q]}\norm{L_cf}^2}^{\frac12}\tag{Cauchy--Schwarz ineq.}\\
    &\le \tuple{2a\ip{f}{Lf}\sum_{c\in[q]}\norm{L_cf}^2}^{\frac12} \tag{\pref{lem:mean}\hyperref[lem:mean-a]{(a)}}\\
    &\le \frac14\ip{f}{Lf}+2a\sum_{c\in[q]}\norm{L_cf}^2\tag{Young's ineq.},
  \end{align*}
  \begin{align*}
    -\sum_{c\in[q]}\+V_c\tuple{f-\widetilde f_c,f}
    &\le \tuple{\sum_{c\in[q]} \+V_c\tuple{f-\widetilde f_c}}^{\frac12}\tuple{\sum_{c\in[q]}\+V_c\tuple{f}}^{\frac12}\tag{Cauchy--Schwarz ineq.}\\
    &\le \tuple{\ip{f}{Lf}\sum_{c\in[q]} \+V_c\tuple{f-\widetilde f_c}}^{\frac12} \tag{\pref{lem:mean-sum}}\\
    &\le \frac18\ip{f}{Lf}+2\sum_{c\in[q]} \+V_c\tuple{f-\widetilde f_c}\tag{Young's ineq.},
  \end{align*}
  \begin{align*}
    \sum_{c\in[q]}\+M_c\tuple{\widetilde f_c,f}
    &\le \tuple{\sum_{c\in[q]} \+M_c\tuple{\widetilde f_c}}^{\frac12}\tuple{\sum_{c\in[q]}\+M_c\tuple{f}}^{\frac12}\tag{Cauchy--Schwarz ineq.}\\
    &\le \frac14\sum_{c\in[q]}\+M_c\tuple{f}+\sum_{c\in[q]} \+M_c\tuple{\widetilde f_c} \tag{Young's ineq.}\\
    &\le \frac14\sum_{c\in[q]}\+M_c\tuple{f}+\frac18\ip{f}{Lf} \tag{\pref{lem:mean}\hyperref[lem:mean-b]{(b)} with $a\theta_\gamma=3$}.
  \end{align*}

  Combining these three inequalities gives
  \begin{equation*}
    \frac34\sum_{c\in[q]}\+M_c\tuple{f}
    \le\frac12\ip f{Lf}+2a\sum_{c\in[q]}\norm{L_cf}^2+2\sum_{c\in[q]}\+V_c\tuple{f-\widetilde f_c}.
  \end{equation*}
  Since $\sum_c\+M_c(f)\ge\ip f{Lf}$ by \pref{lem:mean-sum}, this implies
  \begin{equation*}
    \sum_{c\in[q]}\norm{L_cf}^2
    \ge\frac1{8a}\ip f{Lf}-\frac1a\sum_{c\in[q]}\+V_c\tuple{f-\widetilde f_c}.
  \end{equation*}
  Finally, $1/(8a)=\theta_\gamma/24$, and by \pref{lem:variance} and $1/(q-\Delta)\le \ee/\Delta$,
  \begin{equation*}
    \frac1a\sum_{c\in[q]}\+V_c\tuple{f-\widetilde f_c}
    \le\frac{C_0\ee}\Delta\tuple{1+\frac{\ee a}\Delta}\tuple{\norm{Lf}^2+\ip f{Lf}}.
  \end{equation*}
  Since $a$ depends only on $\gamma$, this proves \pref{eq:hardcore-comparison}.
\end{proof}

\section{Resolvent estimates}\label{sec:resolvent}

In this section, we prove \pref{lem:mean} and \pref{lem:variance}.
Throughout, $a>0$ is fixed, and $R_\eta=(\Id+aH_\eta)^{-1}$, $\widehat R_c$, and $\widetilde f_c$ are as in \pref{sec:two-bounds}.
We first record elementary properties of the resolvent of the Glauber dynamics for a hard-core model,
which we then apply to the hidden model $\nu_\eta$ for each background coloring $\eta$.
Together with \pref{lem:compression}, they readily give \pref{lem:mean} in \pref{sec:mean-proof}.
The rest of the section is devoted to \pref{lem:variance}, whose proof we outline at the beginning of \pref{sec:background-recolorings}.

\begin{fact}\label{fact:resolvent}
  Let $\nu$ and $H$ be as in \pref{sec:prelim-hardcore}, let $a>0$, and let $R\defeq(\Id+aH)^{-1}$.
  For every $h\in L^2(\nu)$, the function $Rh$ is the unique minimizer of
  \begin{equation*}
    \norm[\nu]{g-h}^2+a\ip[\nu]g{Hg}
    \qquad\text{over }g\in L^2(\nu).
  \end{equation*}
  Moreover,
  \begin{equation*}
    \norm[\nu]{Rh}\le\norm[\nu]h,\qquad
    \norm[\nu]{(\Id-R)h}^2\le a\ip[\nu]h{Hh},\qquad
    \ip[\nu]{Rh}{HRh}\le\frac{\ip[\nu]h{Hh}}{(1+a\gap(H))^2}.
  \end{equation*}
  Finally, if $H'$ and $R'\defeq(\Id+aH')^{-1}$ are the operators corresponding to another choice of activities $\lambda'$,
  then $R'-R=aR'(H-H')R$ as linear maps on $\^R^{\+I}$.
\end{fact}

\begin{proof}
  Since $H=n(\Id-P^\nu)$ is self-adjoint and positive semidefinite on $L^2(\nu)$, the objective is strictly convex,
  and its unique minimizer satisfies $(\Id+aH)g=h$, so it equals $Rh$.
  Comparing the objective value at this minimizer with that at $g=h$ gives
  \begin{equation*}
    \norm[\nu]{Rh-h}^2+a\ip[\nu]{Rh}{HRh}\le a\ip[\nu]h{Hh}.
  \end{equation*}
  Discarding the nonnegative second term proves the second inequality.

  For the first and last inequalities, note that the smallest eigenvalue of $H$ is $0$, attained by the constant functions, and that all other eigenvalues,
  counted with multiplicity, are at least $n(1-\lambda_2(P^\nu))=\gap(H)$.
  The operator $R$ multiplies each eigenfunction of $H$ with eigenvalue $z$ by $1/(1+az)$.
  Expanding $h$ in an orthonormal eigenbasis of $H$ therefore reduces these two inequalities to
  \begin{equation*}
    \frac1{(1+az)^2}\le1,\qquad\frac z{(1+az)^2}\le\frac z{(1+a\gap(H))^2},
  \end{equation*}
  which hold for $z=0$ and for every $z\ge\gap(H)$.

  For the resolvent identity, let $\nu'$ be the hard-core model with activities $\lambda'$.
  Since $H$ and $H'$ are positive semidefinite on $L^2(\nu)$ and $L^2(\nu')$, respectively, the operators $\Id+aH$ and $\Id+aH'$ are invertible on $\^R^{\+I}$.
  Multiplying $(\Id+aH)-(\Id+aH')=a(H-H')$ by $R'$ on the left and by $R$ on the right gives the identity.
\end{proof}

\subsection{Displacement and smoothed mean parts: Proof of \texorpdfstring{\pref{lem:mean}}{Lemma \getrefnumber{lem:mean}}}\label{sec:mean-proof}

By \pref{lem:lift}, conditionally on $\eta$, the independent set $I$ has distribution $\nu_\eta$.
Hence, by the law of total expectation, every $F\in L^2(\muh_c)$ satisfies
\begin{equation}\label{eq:conditional-norm}
  \norm[\muh_c]F^2=\E[\eta\sim\rho_c]{\norm[\nu_\eta]{F_\eta}^2},
\end{equation}
where, as in \pref{sec:surviving}, $\rho_c$ denotes the marginal distribution of the background coloring.

\begin{proof}[Proof of \pref{lem:mean}]
  (a) Fix $c\in[q]$.
  Since $\down{(\lift f)}=f$ and the averaging operator is linear, we have $f-\widetilde f_c=\down{((\Id-\widehat R_c)\lift f)}$,
  and $((\Id-\widehat R_c)\lift f)_\eta=(\Id-R_\eta)\lift f_\eta$ for every $\eta$ by the definition of $\widehat R_c$.
  Therefore, using \pref{lem:star} in the last step, we obtain
  \begin{align*}
    \sum_{c\in[q]}\norm{f-\widetilde f_c}^2
    &\overset{\pref{eq:averaging-contraction}}{\le}\sum_{c\in[q]}\norm[\muh_c]{\tuple{\Id-\widehat R_c}\lift f}^2
    \overset{\pref{eq:conditional-norm}}{=}\sum_{c\in[q]}\E[\eta\sim\rho_c]{\norm[\nu_\eta]{\tuple{\Id-R_\eta}\lift f_\eta}^2}\\
    &\overset{\text{\pref{fact:resolvent}}}{\le}a\sum_{c\in[q]}\E[\eta\sim\rho_c]{\ip[\nu_\eta]{\lift f_\eta}{H_\eta\lift f_\eta}}
    \overset{\pref{eq:lift-expectation}}{=}a\sum_{c\in[q]}\ip f{L_cf}
    =2a\ip f{Lf}.
  \end{align*}

  (b) Fix $c\in[q]$ and let $F\defeq\widehat R_c\lift f$, so that $\down F=\widetilde f_c$ and $F_\eta=R_\eta\lift f_\eta$ for every $\eta\in\Omega^{-c}$.
  Since $\gap(H_\eta)\ge\theta$ by assumption, the last inequality of \pref{fact:resolvent}, applied to the hidden model $\nu_\eta$, gives
  \begin{equation*}
    \ip[\nu_\eta]{F_\eta}{H_\eta F_\eta}\le\frac1{\tuple{1+a\theta}^2}\ip[\nu_\eta]{\lift f_\eta}{H_\eta\lift f_\eta}.
  \end{equation*}
  Combining this bound with both parts of \pref{lem:compression} and summing over $c$ using \pref{lem:star}, we obtain
  \begin{align*}
    \sum_{c\in[q]}\+M_c\tuple{\widetilde f_c}
    &\overset{\pref{eq:compression-b}}{\le}\sum_{c\in[q]}\E[\eta\sim\rho_c]{\ip[\nu_\eta]{F_\eta}{H_\eta F_\eta}}
    \le\frac1{\tuple{1+a\theta}^2}\sum_{c\in[q]}\E[\eta\sim\rho_c]{\ip[\nu_\eta]{\lift f_\eta}{H_\eta\lift f_\eta}}\\
    &\overset{\pref{eq:lift-expectation}}{=}\frac1{\tuple{1+a\theta}^2}\sum_{c\in[q]}\ip f{L_cf}
    =\frac2{\tuple{1+a\theta}^2}\ip f{Lf}.\qedhere
  \end{align*}
\end{proof}

\subsection{Background recolorings}\label{sec:background-recolorings}

We now turn to \pref{lem:variance}.
First, we bound $\+V_c$ by a Dirichlet form $\+W_c$ that records the effect of recoloring the background at a vertex where a birth can occur (\pref{lem:variance-pairs}).
Since $f-\widetilde f_c=\down{((\Id-\widehat R_c)\lift f)}$,
it then suffices to compare $(\Id-R_\eta)\lift f_\eta$ with $(\Id-R_{\eta'})\lift f_{\eta'}$ for background colorings $\eta$ and $\eta'$ that differ at a single vertex $v$.
By the resolvent identity in \pref{fact:resolvent}, this difference has one contribution from the change of the activities,
which occurs only at the neighbors of $v$ (\pref{sec:background}), and another from the change of the lifted function,
which is controlled by the mixed second differences of $f$ (\pref{sec:rect-mixed}).
We combine these estimates in \pref{sec:variance}.

Throughout the rest of this section, we assume that $q\ge(1+1/\ee)\Delta$, as in \pref{lem:variance}, in addition to the standing assumption $q-\Delta\ge3$.
We make no attempt to optimize the numerical constants, since each error term only needs to be of order $O(1/\Delta)$ relative to $\norm{Lf}^2+\ip f{Lf}$.
Define
\begin{equation*}
  K\defeq\exp\tuple{\frac\Delta{q-\Delta-1}}.
\end{equation*}
Then, for every coloring $\tau\in\Omega$ and every vertex $v\in V$, we have
\begin{equation}\label{eq:crude}
  \begin{gathered}
    q-\Delta\le|A_v\tuple{\tau}|\le q,\qquad
    q-\Delta-1\ge\frac23\tuple{q-\Delta},\\
    \frac\Delta{q-\Delta}\le \ee,\qquad
    \frac q{q-\Delta}\le1+\ee,\qquad
    K\le \ee^{3\ee/2}<60.
  \end{gathered}
\end{equation}
Indeed, the first bound holds because the neighbors of $v$ use at most $\Delta$ colors, the second follows from $q-\Delta\ge3$,
and the last three follow from $q-\Delta\ge\Delta/\ee$ together with the second bound.

Fix a color $c$.
Consider the auxiliary Markov chain on $\Omega^{-c}\times\+I$ that moves from the current pair $(\eta,I)$ as follows:
\begin{enumerate}[label=\textup{(\arabic*)}]
  \item pick a vertex $v\in V$ uniformly at random;
  \item pick a color $c'\in A^{-c}_v(\eta)$ uniformly at random;
  \item if $v\in V_{\mathrm{ubl}}(I)\setminus I$, move to $(\eta^{v\gets c'},I)$ with probability $p_v(\eta)$ and stay at $(\eta,I)$ with probability $1-p_v(\eta)$;
    otherwise, stay at $(\eta,I)$.
\end{enumerate}
Thus the chain keeps $I$ fixed and recolors the background at $v$ only if $I$ contains neither $v$ nor any neighbor of $v$, i.e.,
only if a birth can occur at $v$.
Since $c'\in A^{-c}_v(\eta)$, the new background coloring $\eta^{v\gets c'}$ again lies in $\Omega^{-c}$.

This chain is reversible with respect to $\muh_c$.
Indeed, let $v\in V_{\mathrm{ubl}}(I)\setminus I$ and $c'\in A^{-c}_v(\eta)$.
Since no vertex of $I$ is adjacent to $v$, recoloring $v$ in the background changes none of the activities $\lambda_w(\eta)$ with $w\in I$,
so $\muh_c(\eta^{v\gets c'},I)=\muh_c(\eta,I)$ by \pref{eq:lift-distribution}.
Moreover, recoloring $v$ changes neither $A^{-c}_v(\eta)$ nor $p_v(\eta)$, and $\eta_v\in A^{-c}_v(\eta^{v\gets c'})$,
so the reverse move has the same probability.
Hence, for each fixed $v$, the weight $\muh_c(\eta,I)\,p_v(\eta)/|A^{-c}_v(\eta)|$ assigned to each triple $(\eta,c',I)$ with $v\in V_{\mathrm{ubl}}(I)\setminus I$ and $c'\in A^{-c}_v(\eta)$ is invariant under the involution
\begin{equation}\label{eq:exchange}
  \tuple{\eta,c',I}\longmapsto\tuple{\eta^{v\gets c'},\eta_v,I},
\end{equation}
which gives detailed balance equations.

For $F\colon\Omega^{-c}\times\+I\to\^R$ and $v\in V$, define
\begin{equation*}
  \+W_{v,c}\tuple{F}\defeq\frac12\E[\tuple{\eta,I}\sim\muh_c]{\one\{v\in V_{\mathrm{ubl}}\tuple{I}\setminus I\}\,p_v\tuple{\eta}\,
    \mathbb{E}_v^{-c}\tuple{\tuple{F\tuple{\eta^{v\gets c'},I}-F\tuple{\eta,I}}^2}},
\end{equation*}
and $\+W_c(F)\defeq\sum_v\+W_{v,c}(F)$.
Here and below, extending the notation of \pref{sec:split}, $\mathbb{E}_v^{-c}$ denotes the expectation over $c'$ chosen uniformly from $A^{-c}_v(\eta)$,
with all other variables held fixed.
Thus $\+W_{v,c}(F)$ is half the expected squared change of $F$ in one step of the chain started from $(\eta,I)\sim\muh_c$, conditioned on picking $v$,
and $\+W_c(F)/n$ is the Dirichlet form of the chain;
in particular, $\+W_c$ is normalized in the same way as $L$ and $H$.
The next lemma relates $\+W_c$ to $\+V_c$ through lifting and averaging.

\begin{lemma}\label{lem:variance-pairs}
  \leavevmode
  \begin{enumerate}[label=\textup{(\alph*)}]
    \item\label{lem:variance-pairs-a} $\+W_{v,c}(\lift f)=\+V_{v,c}(f)$ for every $f\in L^2(\mu)$ and every vertex $v$.
    \item\label{lem:variance-pairs-b} $\+V_c(\down F)\le \+W_c(F)$ for every function $F\colon\Omega^{-c}\times\+I\to\^R$.
  \end{enumerate}
\end{lemma}

\begin{proof}
  Fix $v$, let $(\eta,I)\sim\muh_c$, and write $\sigma\defeq\pi_c(\eta,I)$, which has distribution $\mu$.
  Since $\eta\in\Omega^{-c}$, the event $v\in V_{\mathrm{ubl}}(I)\setminus I$ coincides with the event that $\sigma_v\ne c$ and $c\in A_v(\sigma)$.
  On this event, $\sigma$ and $\eta$ agree at $v$ and on $N(v)$, so $A_v(\eta)=A_v(\sigma)$ and $p_v(\eta)=p_v(\sigma)$.
  Moreover, for every $c'\in A^{-c}_v(\sigma)$, we have $\pi_c(\eta^{v\gets c'},I)=\sigma^{v\gets c'}$,
  and recoloring $v$ with $c'$ leaves the colors available at the vertices of $I$ unchanged, since none of them is adjacent to $v$.
  Hence, conditionally on $\pi_c(\eta,I)=\sigma$, the pair $(\eta^{v\gets c'},I)$ is a random lift of $\sigma^{v\gets c'}$.

  (a) Take $F=\lift f$.
  Then,
  \begin{equation}\label{eq:W-leaf}
    \+W_{v,c}\tuple{\lift f}=\frac12\E[\sigma\sim\mu]{\one\{\sigma_v\ne c,\ c\in A_v\tuple{\sigma}\}\,p_v\tuple{\sigma}\,
      \mathbb{E}_v^{-c}\tuple{\tuple{f\tuple{\sigma^{v\gets c'}}-f\tuple{\sigma}}^2}},
  \end{equation}
  where $\mathbb{E}_v^{-c}$ denotes the expectation over $c'$ chosen uniformly from $A^{-c}_v(\sigma)$.
  We compute the conditional expectation given $\sigma_{V\setminus\{v\}}$.
  If $c\notin A_v(\sigma)$, neither \pref{eq:W-leaf} nor $\+V_{v,c}(f)$ receives a contribution.
  Otherwise, $\sigma_v$ is uniform on $A_v(\sigma)$, so it equals $c$ with probability $p_v(\sigma)$ and lies in $A^{-c}_v(\sigma)$ with probability $1-p_v(\sigma)$;
  in the latter case, $\sigma_v$ and $c'$ are independent and uniform on $A^{-c}_v(\sigma)$.
  Since half the expected squared difference between the values of $f$ at two independent uniform recolorings of $v$ from $A^{-c}_v(\sigma)$ equals $\operatorname{Var}_v^{-c}f(\sigma)$,
  the conditional expectation of the right-hand side of \pref{eq:W-leaf} is
  \begin{equation*}
    \tuple{1-p_v\tuple{\sigma}}\,p_v\tuple{\sigma}\operatorname{Var}_v^{-c}f\tuple{\sigma}
    =\E[\sigma\sim\mu]{\one\{\sigma_v=c\}\tuple{1-p_v\tuple{\sigma}}\operatorname{Var}_v^{-c}f\tuple{\sigma}\;\middle|\;\sigma_{V\setminus\{v\}}}.
  \end{equation*}
  Taking the expectation over $\sigma_{V\setminus\{v\}}$ proves \hyperref[lem:variance-pairs-a]{(a)}.

  (b) Let $F\colon\Omega^{-c}\times\+I\to\^R$, and suppose that $\sigma_v\ne c$ and $c\in A_v(\sigma)$.
  By the definition of $\down F$, for every $c'\in A^{-c}_v(\sigma)$,
  the values $\down F(\sigma^{v\gets c'})$ and $\down F(\sigma)$ are the conditional expectations of $F(\eta^{v\gets c'},I)$ and $F(\eta,I)$, respectively,
  given $\pi_c(\eta,I)=\sigma$.
  Hence, by Jensen's inequality,
  \begin{equation*}
    \tuple{\down F\tuple{\sigma^{v\gets c'}}-\down F\tuple{\sigma}}^2
    \le\E[\tuple{\eta,I}\sim\muh_c]{\tuple{F\tuple{\eta^{v\gets c'},I}-F\tuple{\eta,I}}^2\;\middle|\;\pi_c\tuple{\eta,I}=\sigma}.
  \end{equation*}
  We multiply both sides by $\frac12\one\{\sigma_v\ne c,\ c\in A_v(\sigma)\}\,p_v(\sigma)$ and take the expectation first over $c'$ chosen uniformly from $A^{-c}_v(\sigma)$ and then over $\sigma\sim\mu$.
  By \pref{eq:W-leaf} applied to $\down F$ and by \hyperref[lem:variance-pairs-a]{(a)}, the left-hand side becomes $\+V_{v,c}(\down F)$.
  Since $p_v(\sigma)=p_v(\eta)$ and $A^{-c}_v(\sigma)=A^{-c}_v(\eta)$ on the event in question, the right-hand side becomes $\+W_{v,c}(F)$.
  Thus $\+V_{v,c}(\down F)\le\+W_{v,c}(F)$, and summing over $v$ gives \hyperref[lem:variance-pairs-b]{(b)}.
\end{proof}

\subsection{Effect on the hidden dynamics}\label{sec:background}

Recoloring the background at a single vertex $v$ leaves the graph, the set $\+I$, and the sets $V_{\mathrm{ubl}}(I)$ unchanged,
so it affects the hidden model only through the activities.
It changes $A_u(\eta)$ only for $u\in N(v)$, and for such $u$ it changes $|A_u(\eta)|\ge q-\Delta$ by at most $1$.
Hence the activities $\lambda_u(\eta)=1/(|A_u(\eta)|-1)$ and the conditional occupation probabilities $p_u(\eta)=\lambda_u(\eta)/(1+\lambda_u(\eta))$ change only on $N(v)$,
and only slightly.
The following lemma quantifies the resulting effect on the hidden model and its dynamics.

\begin{lemma}\label{lem:background}
  Fix $c\in[q]$, $\eta\in\Omega^{-c}$, $v\in V$, and $c'\in A^{-c}_v(\eta)$, and set $\eta'\defeq\eta^{v\gets c'}$.
  Then the following hold.
  \begin{enumerate}[label=\textup{(\alph*)}]
    \item\label{lem:background-a} $K^{-1}\nu_\eta(I)\le\nu_{\eta'}(I)\le K\nu_\eta(I)$ for every independent set $I$.
    \item\label{lem:background-b} $\norm[\nu_\eta]{(H_\eta-H_{\eta'})h}^2\le\dfrac{4\Delta}{(q-\Delta)^3}\sum_{u\in N(v)}\ip[\nu_\eta]h{H_{\eta,u}h}$ for every function $h\colon\+I\to\^R$.
  \end{enumerate}
\end{lemma}

\begin{proof}
  (a) For each $u\in N(v)$, the activity ratio $r_u\defeq\lambda_u(\eta')/\lambda_u(\eta)=(|A_u(\eta)|-1)/(|A_u(\eta')|-1)$ lies between $(1+1/(q-\Delta-1))^{-1}$ and $1+1/(q-\Delta-1)$,
  while $\lambda_u(\eta')=\lambda_u(\eta)$ for every $u\notin N(v)$.
  By \pref{eq:hardcore}, $\nu_{\eta'}(I)/\nu_\eta(I)=\zeta(I)/\E[\nu_\eta]{\zeta}$, where $\zeta(I)\defeq\prod_{u\in I\cap N(v)}r_u$.
  For every $I$, each neighbor $u$ contributes a factor of either $1$ or $r_u$, so
  \begin{equation*}
    \prod_{u\in N\tuple{v}}\min\{1,r_u\}\le\zeta\tuple{I}\le\prod_{u\in N\tuple{v}}\max\{1,r_u\}.
  \end{equation*}
  The ratio between the upper and lower bounds is $\prod_{u\in N(v)}\max\{r_u,1/r_u\}\le(1+1/(q-\Delta-1))^\Delta\le K$.
  Since the normalizing factor $\E[\nu_\eta]{\zeta}$ lies between the same bounds, dividing by it gives $K^{-1}\le\nu_{\eta'}(I)/\nu_\eta(I)\le K$,
  which proves \hyperref[lem:background-a]{(a)}.

  (b) Under $\nu_\eta$ and $\nu_{\eta'}$, an unblocked vertex $u$ is occupied with conditional probability $p_u(\eta)$ and $p_u(\eta')$, respectively,
  and these agree for $u\notin N(v)$.
  Subtracting the two instances of \pref{eq:hc-generator} gives, for every independent set $I$,
  \begin{equation*}
    \tuple{\tuple{H_\eta-H_{\eta'}}h}\tuple{I}
    =\sum_{u\in N\tuple{v}}\one\{u\in V_{\mathrm{ubl}}\tuple{I}\}\tuple{p_u\tuple{\eta'}-p_u\tuple{\eta}}\tuple{h\tuple{I\cup\{u\}}-h\tuple{I\setminus\{u\}}}.
  \end{equation*}
  Since $|A_u(\eta)|$ and $|A_u(\eta')|$ differ by at most $1$, we have $|p_u(\eta')-p_u(\eta)|\le p_u(\eta)\,p_u(\eta')$.
  Moreover, $p_u(\eta')\le1/(q-\Delta)$ and $p_u(\eta)/(1-p_u(\eta))=\lambda_u(\eta)\le1/(q-\Delta-1)$, so
  \begin{equation*}
    \tuple{p_u\tuple{\eta'}-p_u\tuple{\eta}}^2\le p_u\tuple{\eta}\tuple{1-p_u\tuple{\eta}}\lambda_u\tuple{\eta}\,p_u\tuple{\eta'}^2
    \le\frac{p_u\tuple{\eta}\tuple{1-p_u\tuple{\eta}}}{\tuple{q-\Delta-1}^3}.
  \end{equation*}
  We apply the Cauchy--Schwarz inequality to the sum over the at most $\Delta$ neighbors $u\in N(v)$, take the expectation under $\nu_\eta$,
  and compare the result with \pref{eq:hc-unblock}.
  Since $\Delta/(q-\Delta-1)^3\le\frac{27}8\Delta/(q-\Delta)^3\le4\Delta/(q-\Delta)^3$ by \pref{eq:crude}, this yields \hyperref[lem:background-b]{(b)}.
\end{proof}

\subsection{Effect on the lifted function}\label{sec:rect-mixed}

A background recoloring also changes the lifted function.
This effect enters the proof of \pref{lem:variance} through
\begin{equation*}
  \Xi\tuple{f}\defeq\sum_{c\in[q]}\sum_{u\in V}\E[\eta\sim\rho_c]{\mathbb{E}_u^{-c}\tuple{
      \ip[\nu_\eta]{\lift f_{\eta^{u\gets c'}}-\lift f_\eta}{H_\eta\tuple{\lift f_{\eta^{u\gets c'}}-\lift f_\eta}}}},
\end{equation*}
the Dirichlet form, under the hidden dynamics, of the change of the lifted function caused by recoloring the background at $u$, summed over $u$ and $c$.
Splitting $H_\eta=\sum_vH_{\eta,v}$ separates the update at $u$ itself from the updates at the other vertices $v$.
The former reduces to the variance part $\+V_{u,c}(f)$.
For the latter, a death at $v$, which recolors $v$ from $c$ to its ghost color, combined with the background recoloring at $u$,
produces a mixed second difference of $f$ over the four colorings obtained by recoloring $u$ and $v$;
these mixed second differences are collected in $\Theta(f)$ (\pref{sec:mixed-squares}).

\begin{lemma}\label{lem:mixed}
  For every $f\in L^2(\mu)$, we have $\Xi(f)\le2\ip f{Lf}+6\Theta(f)$.
  In particular,
  \begin{equation*}
    \Xi\tuple{f}\le96\norm{Lf}^2+622\ip f{Lf}.
  \end{equation*}
\end{lemma}

\begin{proof}
  Fix $c$ and $u$.
  Let $(\eta,I)\sim\muh_c$, so that $I\sim\nu_\eta$ conditionally on $\eta$ by \pref{lem:lift};
  let $c'$ be uniform on $A^{-c}_u(\eta)$ given $(\eta,I)$;
  and write $\eta'\defeq\eta^{u\gets c'}$ and $h\defeq\lift f_{\eta'}-\lift f_\eta$.
  All expectations below are taken jointly over $\eta$, $I$, and $c'$, so that $\E{\ip[\nu_\eta]h{H_\eta h}}$ is the term of $\Xi(f)$ indexed by $c$ and $u$.
  We split $\ip[\nu_\eta]h{H_\eta h}=\sum_v\ip[\nu_\eta]h{H_{\eta,v}h}$.
  Note that $h$ vanishes on every independent set containing $u$, since $\pi_c(\eta',I)=\pi_c(\eta,I)$ whenever $u\in I$.

  We first consider the update at $u$.
  Since $h(I\cup\{u\})=0$, the birth expression \pref{eq:hc-birth} gives
  \begin{equation*}
    \ip[\nu_\eta]h{H_{\eta,u}h}
    =\E[I\sim\nu_\eta]{\one\{u\in V_{\mathrm{ubl}}\tuple{I}\setminus I\}\,p_u\tuple{\eta}h\tuple{I}^2}.
  \end{equation*}
  Since $h(I)=\lift f(\eta^{u\gets c'},I)-\lift f(\eta,I)$, taking the expectation over $\eta$ and $c'$ and comparing with the definition of $\+W_{u,c}$,
  we obtain from \pref{lem:variance-pairs}\hyperref[lem:variance-pairs-a]{(a)} that
  \begin{equation*}
    \E{\ip[\nu_\eta]h{H_{\eta,u}h}}=2\+W_{u,c}\tuple{\lift f}=2\+V_{u,c}\tuple{f}.
  \end{equation*}
  Summed over $c$ and $u$, these terms contribute $2\sum_c\+V_c(f)\le2\ip f{Lf}$ by \pref{lem:mean-sum}.

  Now let $v\ne u$.
  By the death expression \pref{eq:hc-death},
  \begin{equation*}
    \E{\ip[\nu_\eta]h{H_{\eta,v}h}}=\E{\one\{v\in I\}\tuple{1-p_v\tuple{\eta}}\tuple{h\tuple{I}-h\tuple{I\setminus\{v\}}}^2},
  \end{equation*}
  and the integrand vanishes unless $v\in I$ and $u\notin I$.
  On this event, set $\sigma\defeq\pi_c(\eta,I)$.
  Then $\sigma_v=c\ne\sigma_u$, and $A_v(\eta)=A_v(\sigma)$ since no neighbor of $v$ lies in $I$;
  in particular, $p_v(\eta)=p_v(\sigma)$.
  For every $c_u\in A^{-c}_u(\eta)$, the colorings
  \begin{equation*}
    \pi_c\tuple{\eta^{u\gets c_u},I}=\sigma^{u\gets c_u},\qquad
    \pi_c\tuple{\eta,I\setminus\{v\}}=\sigma^{v\gets\eta_v},\qquad
    \pi_c\tuple{\eta^{u\gets c_u},I\setminus\{v\}}=\sigma^{u\gets c_u,\,v\gets\eta_v}
  \end{equation*}
  are proper, since they are images of pairs under $\pi_c$.
  In particular, $c_u\in A_u(\sigma)$ and $\sigma^{u\gets c_u,\,v\gets\eta_v}\in\Omega$.
  For new colors $c_u$ at $u$ and $c_v$ at $v$, denote the mixed second difference of $f$ by
  \begin{equation*}
    M_\sigma\tuple{c_u,c_v}\defeq f\tuple{\sigma}-f\tuple{\sigma^{u\gets c_u}}-f\tuple{\sigma^{v\gets c_v}}+f\tuple{\sigma^{u\gets c_u,\,v\gets c_v}},
  \end{equation*}
  so that, taking $c_u=c'$, we have $h(I)-h(I\setminus\{v\})=-M_\sigma(c',\eta_v)$.

  It remains to take the expectation over $c'$ and the ghost color $\eta_v$.
  When $u\sim v$, the set $A^{-c}_u(\eta)$ from which $c'$ is sampled depends on $\eta_v$.
  We therefore first enlarge it to the set of $c_u\in A_u(\sigma)$ with $\sigma^{u\gets c_u,\,v\gets\eta_v}\in\Omega$,
  which contains it by the previous paragraph and depends only on $\sigma$ and $\eta_v$.
  Since $1/|A^{-c}_u(\eta)|=1/(|A_u(\eta)|-1)\le1/(q-\Delta-1)\le q\,p_u(\sigma)/(q-\Delta-1)$ and all terms are nonnegative, we obtain
  \begin{equation*}
    \mathbb{E}_u^{-c}\tuple{M_\sigma\tuple{c',\eta_v}^2}
    \le\frac q{q-\Delta-1}\,p_u\tuple{\sigma}
    \sum_{\substack{c_u\in A_u\tuple{\sigma}\\\sigma^{u\gets c_u,\,v\gets\eta_v}\in\Omega}}M_\sigma\tuple{c_u,\eta_v}^2.
  \end{equation*}
  For fixed $\sigma$, the right-hand side depends on the ghost colors only through $\eta_v$, which is uniform on $A^{-c}_v(\sigma)$ conditionally on $\sigma$.
  Since $1-p_v(\sigma)=p_v(\sigma)\,|A^{-c}_v(\sigma)|$, taking the expectation over $\eta_v$ turns the factor $1-p_v(\sigma)$ into $p_v(\sigma)$ times a sum over $c_v\in A^{-c}_v(\sigma)$.
  We then drop the condition $\sigma_u\ne c$ and add the vanishing terms with $c_v=\sigma_v$;
  neither step decreases the right-hand side.
  This gives
  \begin{equation*}
    \E{\ip[\nu_\eta]h{H_{\eta,v}h}}
    \le\frac q{q-\Delta-1}\E[\sigma\sim\mu]{\one\{\sigma_v=c\}\,p_u\tuple{\sigma}\,p_v\tuple{\sigma}
      \sum_{\substack{c_u\in A_u\tuple{\sigma},\ c_v\in A_v\tuple{\sigma}\\\sigma^{u\gets c_u,\,v\gets c_v}\in\Omega}}M_\sigma\tuple{c_u,c_v}^2}.
  \end{equation*}
  Summing over $c\in[q]$ removes the indicator, and since $\omega_{uv}(\sigma)=\mu(\sigma)\,p_u(\sigma)\,p_v(\sigma)$,
  the definition of $\Theta_{uv}(f)$ in \pref{sec:mixed-squares} gives
  \begin{equation*}
    \sum_{c\in[q]}\E{\ip[\nu_\eta]h{H_{\eta,v}h}}\le\frac q{q-\Delta-1}\Theta_{uv}\tuple{f}\le6\Theta_{uv}\tuple{f},
  \end{equation*}
  where the last inequality uses $q/(q-\Delta-1)\le\frac32\cdot q/(q-\Delta)\le\frac32(1+\ee)$ by \pref{eq:crude}.

  Summing both kinds of terms over $c$, $u$, and $v$ gives $\Xi(f)\le2\ip f{Lf}+6\Theta(f)$.
  The second claim then follows from \pref{lem:rect} and $\Delta/(q-\Delta)\le\ee$, since $6\cdot16=96$ and $2+6\cdot38\ee\le622$.
\end{proof}

\subsection{Variance part of the displacement: Proof of \texorpdfstring{\pref{lem:variance}}{Lemma \getrefnumber{lem:variance}}}\label{sec:variance}

\begin{proof}[Proof of \pref{lem:variance}]
  Fix $c\in[q]$.
  As in the proof of \pref{lem:mean}\hyperref[lem:mean-a]{(a)}, we have $f-\widetilde f_c=\down{((\Id-\widehat R_c)\lift f)}$,
  so \pref{lem:variance-pairs}\hyperref[lem:variance-pairs-b]{(b)} gives
  \begin{equation*}
    \+V_c\tuple{f-\widetilde f_c}
    \le\+W_c\tuple{\tuple{\Id-\widehat R_c}\lift f}=\sum_{v\in V}\+W_{v,c}\tuple{\tuple{\Id-\widehat R_c}\lift f}.
  \end{equation*}
  Now fix also $v\in V$.
  For $\eta\in\Omega^{-c}$ and $c'\in A^{-c}_v(\eta)$, write $\eta'\defeq\eta^{v\gets c'}$.
  The resolvent identity in \pref{fact:resolvent}, applied to the activities $\lambda(\eta)$ and $\lambda(\eta')$,
  gives $R_{\eta'}-R_\eta=aR_{\eta'}(H_\eta-H_{\eta'})R_\eta$, and hence
  \begin{equation}\label{eq:commute}
    \tuple{\Id-R_{\eta'}}\lift f_{\eta'}-\tuple{\Id-R_\eta}\lift f_\eta
    =\tuple{\Id-R_{\eta'}}\tuple{\lift f_{\eta'}-\lift f_\eta}-aR_{\eta'}\tuple{H_\eta-H_{\eta'}}R_\eta\lift f_\eta.
  \end{equation}
  Substituting \pref{eq:commute} into the definition of $\+W_{v,c}((\Id-\widehat R_c)\lift f)$ and using $(s-t)^2\le2s^2+2t^2$, we obtain
  \begin{equation*}
    \begin{aligned}
      &\+W_{v,c}\tuple{\tuple{\Id-\widehat R_c}\lift f}\\
      &\quad\le\E[\tuple{\eta,I}\sim\muh_c]{\one\{v\in V_{\mathrm{ubl}}\tuple{I}\setminus I\}\,p_v\tuple{\eta}\,
        \mathbb{E}_v^{-c}\tuple{\tuple{\tuple{\Id-R_{\eta'}}\tuple{\lift f_{\eta'}-\lift f_\eta}}\tuple{I}^2}}\\
      &\qquad+a^2\E[\tuple{\eta,I}\sim\muh_c]{\one\{v\in V_{\mathrm{ubl}}\tuple{I}\setminus I\}\,p_v\tuple{\eta}\,
        \mathbb{E}_v^{-c}\tuple{\tuple{R_{\eta'}\tuple{H_\eta-H_{\eta'}}R_\eta\lift f_\eta}\tuple{I}^2}}.
    \end{aligned}
  \end{equation*}

  We bound the two terms separately.
  In the first term, we apply the involution \pref{eq:exchange} \emph{before} dropping the indicator.
  The involution preserves the full weight, swaps $\eta$ and $\eta'$, and changes the sign of $\lift f_{\eta'}-\lift f_\eta$;
  since the sign disappears after squaring, this replaces $R_{\eta'}$ by $R_\eta$ without changing the first term.
  The indicator is essential because, outside its event, recoloring $v$ may change the activity of an occupied neighbor,
  and then the weight need not be preserved.
  We then drop the indicator, bound $p_v(\eta)\le1/(q-\Delta)$, and take the expectation over $I\sim\nu_\eta$ with $\eta$ fixed.
  Together with the second inequality of \pref{fact:resolvent}, this bounds the first term by
  \begin{equation*}
    \begin{aligned}
      &\frac1{q-\Delta}\E[\eta\sim\rho_c]{\mathbb{E}_v^{-c}\tuple{
          \norm[\nu_\eta]{\tuple{\Id-R_\eta}\tuple{\lift f_{\eta'}-\lift f_\eta}}^2}}\\
      &\qquad\le\frac a{q-\Delta}\E[\eta\sim\rho_c]{\mathbb{E}_v^{-c}\tuple{
          \ip[\nu_\eta]{\lift f_{\eta'}-\lift f_\eta}{H_\eta\tuple{\lift f_{\eta'}-\lift f_\eta}}}}.
    \end{aligned}
  \end{equation*}
  By the definition of $\Xi(f)$, these bounds sum over $c$ and $v$ to $a\,\Xi(f)/(q-\Delta)$.

  In the second term, we drop the indicator directly, bound $p_v(\eta)\le1/(q-\Delta)$, and take the expectation over $I\sim\nu_\eta$ with $\eta$ fixed.
  This bounds the second term by
  \begin{equation*}
    \frac{a^2}{q-\Delta}\E[\eta\sim\rho_c]{\mathbb{E}_v^{-c}\tuple{
        \norm[\nu_\eta]{R_{\eta'}\tuple{H_\eta-H_{\eta'}}R_\eta\lift f_\eta}^2}}.
  \end{equation*}
  To estimate this norm, we compare the norms under $\nu_\eta$ and $\nu_{\eta'}$ by \pref{lem:background}\hyperref[lem:background-a]{(a)},
  apply the first inequality of \pref{fact:resolvent} to $R_{\eta'}$, compare the norms again,
  and finally apply \pref{lem:background}\hyperref[lem:background-b]{(b)} to obtain
  \begin{equation*}
    \begin{aligned}
      \norm[\nu_\eta]{R_{\eta'}\tuple{H_\eta-H_{\eta'}}R_\eta\lift f_\eta}^2
      &\le K\norm[\nu_{\eta'}]{R_{\eta'}\tuple{H_\eta-H_{\eta'}}R_\eta\lift f_\eta}^2
      \le K\norm[\nu_{\eta'}]{\tuple{H_\eta-H_{\eta'}}R_\eta\lift f_\eta}^2\\
      &\le K^2\norm[\nu_\eta]{\tuple{H_\eta-H_{\eta'}}R_\eta\lift f_\eta}^2
      \le\frac{4K^2\Delta}{\tuple{q-\Delta}^3}\sum_{u\in N\tuple{v}}\ip[\nu_\eta]{R_\eta\lift f_\eta}{H_{\eta,u}R_\eta\lift f_\eta}.
    \end{aligned}
  \end{equation*}
  The function $R_\eta\lift f_\eta$ depends neither on $v$ nor on $c'$.
  Since every vertex $u$ has at most $\Delta$ neighbors $v$ and each $H_{\eta,u}$ is positive semidefinite, the last inequality of \pref{fact:resolvent} gives
  \begin{equation*}
    \sum_{v\in V}\sum_{u\in N\tuple{v}}\ip[\nu_\eta]{R_\eta\lift f_\eta}{H_{\eta,u}R_\eta\lift f_\eta}
    \le\Delta\ip[\nu_\eta]{R_\eta\lift f_\eta}{H_\eta R_\eta\lift f_\eta}
    \le\Delta\ip[\nu_\eta]{\lift f_\eta}{H_\eta\lift f_\eta}.
  \end{equation*}
  Taking the expectation over $\eta$ and summing over $c$, we obtain from \pref{lem:compression}\hyperref[lem:compression-a]{(a)} and \pref{lem:star} that
  \begin{equation*}
    \sum_{c\in[q]}\E[\eta\sim\rho_c]{\ip[\nu_\eta]{\lift f_\eta}{H_\eta\lift f_\eta}}
    =\sum_{c\in[q]}\ip f{L_cf}=2\ip f{Lf}.
  \end{equation*}
  Altogether, using $\Delta/(q-\Delta)\le\ee$ from \pref{eq:crude}, we find that the bounds for the second term sum over $c$ and $v$ to at most
  \begin{equation*}
    \frac{a^2}{q-\Delta}\cdot\frac{4K^2\Delta}{\tuple{q-\Delta}^3}\cdot\Delta\cdot2\ip f{Lf}
    \le\frac{8\ee^2K^2a^2}{\tuple{q-\Delta}^2}\ip f{Lf}.
  \end{equation*}

  Combining the bounds for the two terms with \pref{lem:mixed}, we obtain
  \begin{equation*}
    \sum_{c\in[q]}\+V_c\tuple{f-\widetilde f_c}
    \le\frac a{q-\Delta}\tuple{96\norm{Lf}^2+622\ip f{Lf}}+\frac{8\ee^2K^2a^2}{\tuple{q-\Delta}^2}\ip f{Lf}.
  \end{equation*}
  Since $K<60$ by \pref{eq:crude}, we have $8\ee^2K^2<2.2\cdot10^5$, so \pref{lem:variance} holds with $C_0=2.2\cdot10^5$.
\end{proof}

\section*{Acknowledgments}

The author thanks Weiming Feng, Heng Guo, Jingcheng Liu, Eric Vigoda, and Xinyuan Zhang for discussions,
and is especially grateful to Xiaoyu Chen for introducing him to recently developed Bochner-type inequalities in approximate counting and sampling and for many insightful discussions during the Local to Global Principles workshop at the Simons Institute for the Theory of Computing.
This work was supported in part by NSF grant CCF-2147094.

\section*{AI Methodology}

The main ideas of this paper emerged from extensive discussions with ChatGPT-6 Astra Pro.
ChatGPT-6 Astra Pro also suggested that a finer analysis of the slack by which the hidden hard-core models lie inside the uniqueness regime should improve the constant $1+1/\ee$ further.
We nevertheless keep to the present setting, which we believe shows the idea of the hidden hard-core model most cleanly.

The author then simplified several arguments in the first version of the proof and wrote the final text, using ChatGPT-6 Astra, Claude Opus 5.5,
and Gemini 4 Argon for editorial suggestions.
All mathematical claims have been independently verified by the author, who takes full responsibility for the content of this paper.

\printbibliography

\end{document}